\documentclass{article}

\usepackage{arxiv}

\usepackage{amsmath}
\usepackage[utf8]{inputenc} % allow utf-8 input
\usepackage[T1]{fontenc}    % use 8-bit T1 fonts
\usepackage{hyperref}       % hyperlinks
\usepackage{url}            % simple URL typesetting
\usepackage{booktabs}       % professional-quality tables
\usepackage{amsfonts}       % blackboard math symbols
\usepackage{nicefrac}       % compact symbols for 1/2, etc.
\usepackage{microtype}      % microtypography
\usepackage{cleveref}       % smart cross-referencing
\usepackage{lipsum}         % Can be removed after putting your text content
\usepackage{graphicx}
\usepackage{natbib}
\usepackage{doi}
\usepackage{multirow}
\usepackage{amsthm,amssymb,amsfonts}
\usepackage{threeparttable}

\title{First Principles based High-precision Modelling and Identification of Piezoelectric Fast Steering Mirror}

\newif\ifuniqueAffiliation
\uniqueAffiliationtrue

\ifuniqueAffiliation % Standard variant of author block
\author{ \href{https://orcid.org/0000-0002-2471-1574}{\includegraphics[scale=0.06]{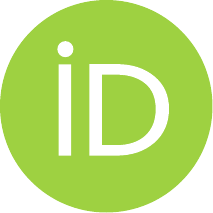}\hspace{1mm}Sen Yang} \\
	School of Astronautics and Aeronautics\\
	University of Electronic Science and Technology of China\\
	Chengdu, China 611731 \\
	\texttt{yang\_jansen@163.com} \\
	\And
	\href{https://orcid.org/0000-0000-0000-0000}{\includegraphics[scale=0.06]{orcid.pdf}\hspace{1mm}Xiaofeng Li} \\
	School of Astronautics and Aeronautics\\
	University of Electronic Science and Technology of China\\
	Chengdu, China 611731 \\
	\texttt{lxf3203433@uestc.edu.cn} \\
}
\else
\usepackage{authblk}

\newbox{\orcid}\sbox{\orcid}{\includegraphics[scale=0.06]{orcid.pdf}} 
\author[1]{%
	\href{https://orcid.org/0000-0002-2471-1574}{\usebox{\orcid}\hspace{1mm}Sen Yang\thanks{\texttt{yang\_jansen@163.com}}}%
}
\author[1]{%
	\href{https://orcid.org/0000-0000-0000-0000}{\usebox{\orcid}\hspace{1mm}Xiaofeng Li\thanks{\texttt{lxf3203433@uestc.edu.cn}}}%
}
\affil[1]{School of Astronautics and Aeronautics, University of Electronic Science and Technology of China, Chengdu, China 611731}
\fi

\renewcommand{\shorttitle}{\textit{arXiv} Template}

\begin{document}
\maketitle

\begin{abstract}
We establish a high-precision composite model for a piezoelectric fast steering mirror (PFSM) using a Hammerstein structure. A novel asymmetric Bouc-Wen model is proposed to describe the nonlinear rate-independent hysteresis, while a dynamic model is derived to represent the linear rate-dependent component. By analyzing the physical process from the displacement of the piezoelectric actuator to the angle of the PFSM, cross-axis coupling is modeled based on first principles. Given the dynamic isolation of each module on different frequency scales, a step-by-step method for model parameter identification is carried out. Finally, experimental results demonstrate that the identified parameters can accurately represent the hysteresis, creep, and mechanical dynamic characteristics of the PFSM. Furthermore, by comparing the outputs of the identified model with the real PFSM under different excitation signals, the effectiveness of the proposed dual-input dual-output composite model is validated.
\end{abstract}

% keywords can be removed

The piezoelectric fast steering mirror (PFSM), driven by piezoelectric ceramics, is utilized to adjust the direction of laser beams \cite{zhou2008design,haber2023data}. It can achieve sub-microradian precision deflection angles and offers advantages such as high control bandwidth, fine resolution, and effective suppression of high-frequency disturbances. PFSMs are widely applied in precision tracking systems for space optical communication \cite{kluk2007advanced,bekkali2022new}, imaging systems \cite{chen2010modeling,sun2021conceptual}, lithography systems \cite{csencsics2017system,zhong2022design}, and biomedical systems \cite{oscar2015pupil,fernandez2022instrument}, playing a crucial role in military and industrial applications.

Although the PFSM possesses many advantages, challenges arise due to the hysteresis nonlinearity and creep characteristics between the input voltage and output displacement of the piezoelectric actuator (PEA), as well as mechanical dynamic and cross-coupling issues between axes \cite{hao2024rate}. These complexities make it difficult to achieve stable control with high precision and bandwidth using model-free control methods like error-driven proportional-integral-derivative. Hence, the employment of model-based controllers is indispensable \cite{lv2023modeling}. However, the aforementioned nonlinearities and coupling factors present significant challenges in effectively and accurately modeling complex PFSM systems.

Due to the crystal polarization effect of the PEA, its elongation displacement does not linearly correlate with the driving voltage, resulting in nonlinear dynamic hysteresis errors between input and output, which reduces overall system performance \cite{lallart2019system}. Common hysteresis modeling methods are based on physical principles and phenomenology. The Jiles-Atherton (J-A) hysteresis model is a physical model based on first principles \cite{jiles1983ferromagnetic}. Ref. \cite{raghunathan2009modeling} has extended the J-A model with additional thermal parameters to describe the temperature-dependent magnetization process in ferromagnetic materials. However, accurate model identification is quite complex and involves various approximation techniques to solve inverse model equations, making it less precise in expressing the stress-magnetization curve than phenomenological models, which describe hysteresis based on the relationship between input and output \cite{suzuki2005comparison}.

Currently, phenomenological models widely represented for PEA hysteresis characteristics include the Preisach model \cite{mayergoyz2003mathematical,xiao2012modeling} and the Prandtl-Ishlinskii model \cite{al2010analytical,al2021adaptive}, which use weighted elementary hysteresis operators to model the rising and falling of the hysteresis curve. Additionally, the Duhem model \cite{oh2005semilinear,ahmed2021duhem} and the Bouc-Wen model \cite{ismail2009hysteresis,zhang2023hysteresis} employ differential equations to model hysteresis, allowing for convenient conversion into state equations for controller design.

Creep in piezoelectric materials is related to the residual polarization caused by the applied voltage. As the operating voltage increases, the residual polarization continues to rise, manifesting as a slow drift in displacement output when the PEA is subjected to a step input, thus affecting the system absolute positioning performance. The creep rate and magnitude largely depend on the piezoelectric material, implying that the mechanical parameters of the piezoelectric material drift over time \cite{lapchuk2011creep}. In typical operational modes of scanning probe microscopy, long-term positioning during sample measurement leads to image distortion due to creep \cite{habibullah202030}. Therefore, analyzing and modeling the creep phenomenon is crucial.

Currently, there are two primary methods for modeling the creep phenomenon. The first is the time-domain logarithmic model \cite{changhai2005hysteresis}, which relies on the selection of time parameters for fitting the model. However, this model fails as time approaches zero or becomes very large because the output becomes unbounded \cite{devasia2007survey}. To overcome these shortcomings, studies \cite{croft2001creep,gu2013motion} have adopted a frequency-domain model, represented by a series of springs and dampers. This approach utilizes the creep model in mechanical dynamics to obtain the low-frequency response of the PEA. The advantage of this frequency-domain linear model is that it can be directly multiplied with other linear transfer functions, resulting in a composite model that comprehensively describes the linear response of the system, thereby simplifying the design of model-based controllers.

Ref. \cite{zhu2015modeling} represents the hysteresis of PFSM using the Bouc-Wen model and describes the dynamics of PFSM using the transfer matrix method for multibody systems. This approach constructs a mathematical model of the PFSM while maintaining low-order system dynamics without needing the global dynamic equations of the system, thus avoiding the computational difficulties associated with high-order matrices. Ref. \cite{liu2019composite} proposes a composite model that includes Bouc-Wen hysteresis, creep dynamics, and electromechanical dynamics, along with a step-by-step model parameters identification method. However, dual-axis cross-coupling causes the input on one axis to induce an output on the other axis, leading to non-negligible positioning errors in precision tracking systems. Due to the lack of consideration for cross-coupling effects, the controllers designed based on the composite models \cite{zhu2015modeling,liu2019composite} still have room for improvement in tracking accuracy.

To address the issue of decreased PFSM accuracy caused by mechanical cross-coupling between axes, Ref. \cite{o2012cross} estimates the cross-coupling coefficients of PFSM in real time by comparing the measured beam position with the predicted position of a non-coupled system model. Ref. \cite{wang2014fast} improves the online regularized extreme learning machine algorithm for decoupled control of multiple input multiple output (MIMO) nonlinear continuous-time systems. Ref. \cite{ito2023modeling} proposes a control method based on iterative computation of a 2×2 Jacobian matrix to explicitly compensate for the cross-coupling motion of dual axes of the PFSM. However, these methods are data-driven and do not involve mathematical modeling of cross-coupling. Additionally, other works aim to mitigate inter-axis coupling through optimized structural design. Ref. \cite{shao2018two} proposes an overall architecture using a two-degree-of-freedom (2DOF) flexible hinge with cross-axis decoupling capability, achieving a low dynamic cross-coupling ratio between 2DOF. Ref. \cite{chang2021development} designs a PFSM with a differential push-pull structure, featuring high stiffness and good decoupling characteristics. Ref. \cite{han2022design} designs a high-precision miniature PFSM guided by an annular flexible hinge, achieving an ultra-low coupling ratio. These studies provide important insights into the mechanical design of PFSM systems. However, they only cascade a linear matrix in the time domain without considering the internal dynamic processes when modeling cross-coupling. Ref. \cite{wang2019comprehensive} constructs a comprehensive model of the PFSM system by analyzing each physical component, where the modeling of cross-axis coupling involves the dynamic process from PEA displacement to PFSM angle. Nevertheless, it describes the hysteresis of the PFSM single-axis angle only using the hysteresis of single PEA displacement, which is not sufficiently rigorous. Since a single axis is driven by two coupled PEAs, the corresponding hysteresis curve could be more complex.

However, due to the lack of exploration into the intrinsic properties and interrelationships of factors such as the coupling of dual-PEA hysteresis and dual-axis electromechanical cross-coupling, the fine modeling of complex PFSM systems has not yet been well resolved. This paper begins from first principles, progressively analyzing the dynamics of each physical component, and establishes a comprehensive dual-input dual-output high-precision model of the PFSM. Specifically, a Hammerstein structure \cite{butcher2016identification} is employed to decompose the nonlinear rate-dependent characteristics. The nonlinear hysteresis is described using an improved rate-independent asymmetric Bouc-Wen model, while the rate-dependent creep and electromechanical dynamics are represented by linear transfer functions. Furthermore, based on the dynamic isolation of each module on different frequency scales, a method for model parameters identification is proposed.

The remainder of this paper is organized as follows. Section 2 provides a brief introduction to the physical structure and operating principle of the PFSM. Section 3 establishes a composite model for the single-axis PFSM. Building on this, Section 4 derives the composite model for the dual-axis PFSM. Section 5 conducts identification experiments on the dynamics of each module and evaluates the identified composite model. Finally, Section 6 presents the conclusions.

\section{Structure of PFSM} \label{sec:2}
\begin{figure}[htpb]
	\centering\includegraphics[width=0.7\columnwidth]{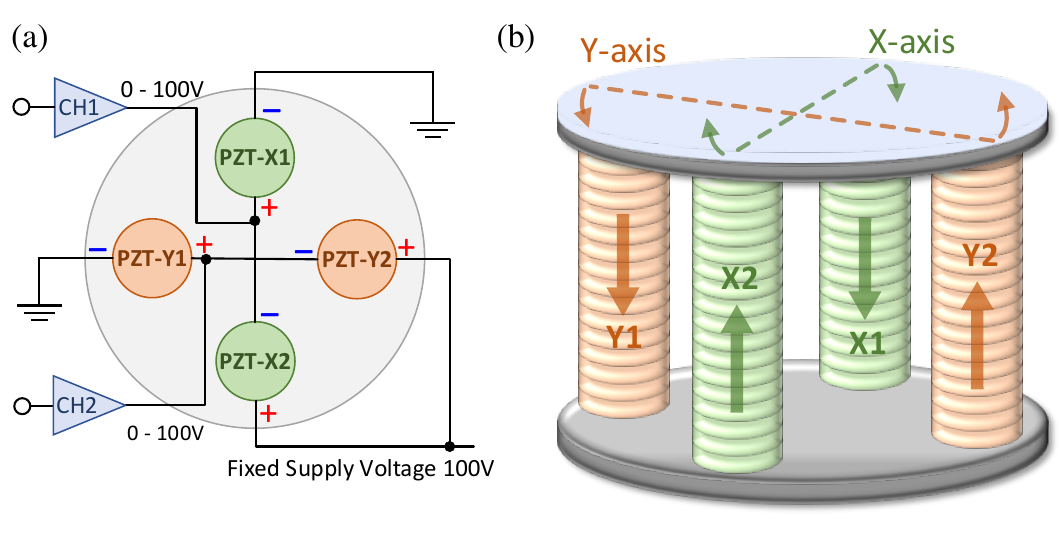}
	\caption{Schematic diagram of a dual-axis PFSM system. (a) Arrangement and circuit of PEAs. (b) Structure of PFSM deflection driven by four PEAs.}
	\label{fig:1}
\end{figure}

The typical structure of a dual-axis PFSM system is shown in Fig. \ref{fig:1}. Generally, its components include two pairs of PEAs, a flexible hinge mechanism, a mirror, and a dual-channel strain gauge sensor (SGS). Fig.\ref{fig:1}(a) displays the relative positions of the four PEAs. For each axis, two PEAs are connected in series, with one grounded and the other connected to a fixed power supply. The grounded PEA expands as the input voltage of the channel increases, while the other contracts. By varying the input voltage signals on the X and Y axes, the two pairs of actuators can drive the PFSM to achieve deflection movements along the X and Y axes, as illustrated in Fig.\ref{fig:1}(b).

It can be observed that the mechanical component of the PFSM is an interconnected integral structure, resulting in some correlation between the outputs of the two axes. We first model the dynamics of a single-axis single-input single-output (SISO) system. Then, by considering the mechanical cross-coupling, we achieve complete dynamic modeling of the dual-axis MIMO system.

\section{Single-axis System Modeling} \label{sec:3}
\begin{figure}[htpb]
	\centering\includegraphics[width=1.0\columnwidth]{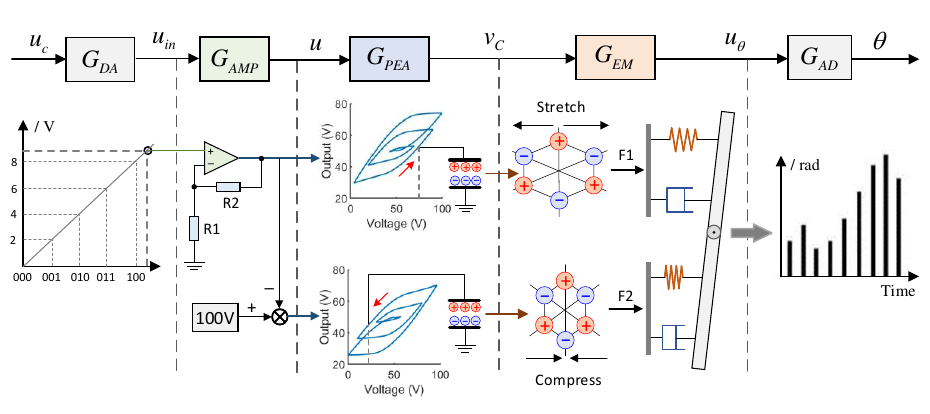}
	\caption{The details and block diagram of the single-axis SISO model.}
	\label{fig:2}
\end{figure}

We first model the dynamics of the single-axis SISO system, as shown in Fig. \ref{fig:2}. By analyzing the physical characteristics of each component, it is decomposed into multiple subsystem modals in series and parallel. The digital controller converts the digital control signal ${{u}_{c}}\left( t \right)$ into an analog signal ${{u}_{in}}\left( t \right)$ through digital-to-analog conversion (DAC) ${{G}_{DA}}$. This analog signal is then amplified by the amplifier module ${{G}_{AMP}}$ to produce the voltage signal $u\left( t \right)$ that drives the platform deflection.

The electrical dynamic of PEA1 is represented by ${{G}_{PEA1}}$, with input ${{u}_{1}}\left( t \right)=u\left( t \right)$ and output being the voltage ${{v}_{C1}}\left( t \right)$ across the equivalent capacitance of actuator. For PEA2, its electrical dynamic is represented by ${{G}_{PEA2}}$, with input ${{u}_{2}}\left( t \right)={{u}_{MAX}}-{{u}_{1}}\left( t \right)$ and output ${{v}_{C2}}\left( t \right)$, where ${{u}_{MAX}}$ is the fixed power supply voltage. The electromechanical component of the PFSM is represented by the transfer function ${{G}_{EM}}$, with inputs ${{v}_{C1}}\left( t \right)$ and ${{v}_{C2}}\left( t \right)$, and the flowing electric charge generates forces ${{F}_{1}}\left( t \right)$ and ${{F}_{2}}\left( t \right)$ acting on the mounting platform. These forces cause structural deformation, which changes the resistance of the strain gauge sensor (SGS). The servo module converts this resistance change into a voltage signal ${{u}_{\theta }}\left( t \right)$. Finally, the data acquisition system performs analog-to-digital conversion (ADC) ${{G}_{AD}}$, providing real-time digital feedback of the deflection angle $\theta$ of PFSM.

It is important to note that the bandwidths of DAC and ADC in the data acquisition system are much greater than the signal bandwidth, so their transfer functions can be simplified to ${{G}_{DA}}={{G}_{AD}}=1$. Similarly, the bandwidth of the amplifier module is also much greater than the signal bandwidth, so it can be considered to linearly amplify the low voltage input ${{u}_{in}}\left( t \right)$ to the driving voltage $u\left( t \right)$, with the transfer function equal to the constant amplification gain, i.e., ${{G}_{AMP}}=1+{{R}_{2}}/{{R}_{1}}={{k}_{AMP}}$.

From the analysis above, it is clear that the PFSM is a complex system with nonlinear and rate-dependent characteristics. In this paper, we model the single-axis system employing a Hammerstein structure, where the nonlinear hysteresis characteristics are described by a rate-independent modified asymmetric Bouc-Wen model, and the rate-dependent dynamic characteristics are represented by a linear transfer function. Below, we will provide a detailed analysis and modeling of the electrical dynamics of the PEA and the mechanical components of the PFSM.

\subsection{The electrical dynamics of the PEA} \label{sec:3.1}
The electrical dynamics ${{G}_{PEA}}$ of the PEA can be represented as a series connection of hysteresis ${{G}_{HYS}}$ and creep ${{G}_{CRP}}$. From Fig. \ref{fig:2}, it can be seen that the input to the PEA is the driving voltage $u\left( t \right)$, and the output is the voltage ${{v}_{C}}\left( t \right)$ across the equivalent capacitance of the actuator.

\subsubsection{Hysteresis Characteristic} \label{sec:3.1.1}
The hysteresis is a typical nonlinear effect of piezoelectric materials. The classical Bouc-Wen hysteresis model is represented as follows \cite{ismail2009hysteresis}:
\begin{equation}
	\left\{ \begin{aligned}
		& {{v}_{h}}(t)=u(t)+h(t) \\ 
		& \dot{h}(t)=\alpha \dot{u}(t)-\beta \left| \dot{u}(t) \right|{{\left| h(t) \right|}^{n-1}}h(t)-\gamma \dot{u}(t){{\left| h(t) \right|}^{n}} \\ 
	\end{aligned} \right.
	\label{eq:1}
\end{equation}
where ${{v}_{h}}\left( t \right)$ is the output after the voltage drop caused by hysteresis nonlinearity. The hysteresis component $h\left( t \right)$ does not exhibit central symmetry during a stable period, as indicated by the larger absolute value of the hysteresis component at higher drive voltages. This phenomenon mainly occurs because the resistance of the PEA to deformation is stronger at higher drive voltages than at lower ones. Consequently, the hysteresis component $h\left( t \right)$ of the PEA changes more rapidly near higher drive voltages, making it asymmetric. To mitigate the sharp increase in modeling error at high drive voltages, an asymmetry factor $\delta u(t)\text{sgn} \left[ \dot{u}(t) \right]$ is introduced into the Bouc-Wen hysteresis operator \cite{zhu2023modeling}:
\begin{equation}
	\left\{ \begin{aligned}
		& {{v}_{h}}(t)=u(t)+h(t) \\ 
		& \dot{h}(t)=\alpha \dot{u}(t)-\beta \left| \dot{u}(t) \right|{{\left| h(t) \right|}^{n-1}}h(t)-\gamma \dot{u}(t){{\left| h(t) \right|}^{n}}+\delta u(t)\text{sgn} \left[ \dot{u}(t) \right] \\ 
	\end{aligned} \right.
	\label{eq:2}
\end{equation}

As the drive voltage increases, the asymmetry factor has an increasing impact on $\dot{h}(t)$. When $\delta $ is negative, the trend of the asymmetry factor changes opposite to that of $\dot{u}(t)$, representing the phenomenon of resistance to deformation. This effectively improves the characterization of the asymmetric properties of the PEA by the Bouc-Wen model.

From Eq. (\ref{eq:1}), it can be seen that $\dot{h}(t)$ is positively correlated with $\dot{u}(t)$. Therefore, with the model parameters determined, $\Delta h$ is only related to $\Delta u$ and is independent of $\Delta t$, meaning that the hysteresis component is related to the driving voltage and is independent of its path. However, the asymmetric factor in method \cite{georgiou2008dynamic} does not include the term $\dot{u}(t)$, which results in the asymmetric model lacking this important rate-independent characteristic. To remedy this deficiency, the asymmetric factor is modified to $\delta \dot{u}(t)u(t)$, and the hysteresis component in Eq. (\ref{eq:2}) is rewritten as:
\begin{equation}
	\left\{ \begin{aligned}
		& {{v}_{h}}(t)=u(t)+h(t) \\ 
		& \dot{h}(t)=\alpha \dot{u}(t)-\beta \left| \dot{u}(t) \right|{{\left| h(t) \right|}^{n-1}}h(t)-\gamma \dot{u}(t){{\left| h(t) \right|}^{n}}+\delta \dot{u}(t)u(t) \\ 
	\end{aligned} \right.
	\label{eq:3}
\end{equation}
where the hysteresis component is rate-independent, inheriting the physical property of the previous asymmetric factor, which exhibits strong resistance to deformation at high driving voltages. Additionally, the improved asymmetric factor enhances the ability to fit the hysteresis nonlinearity. Consequently, our improved asymmetric Bouc-Wen model can more accurately describe the hysteresis characteristics of PEA. The nonlinear transfer function from the driving voltage $u\left( t \right)$ to the hysteresis voltage ${{v}_{h}}\left( t \right)$ is denoted as ${{G}_{HYS}}$:
\begin{equation}
	{{G}_{HYS}}(s)=\frac{{{V}_{H}}(s)}{U(s)}
	\label{eq:4}
\end{equation}
where ${{V}_{H}}\left( s \right)$ and $U\left( s \right)$ are the Laplace transforms of ${{v}_{h}}\left( t \right)$ and $u\left( t \right)$, respectively.

\subsubsection{Creep Dynamic} \label{sec:3.1.2}
Creep is a slow dynamic characteristic that describes the change in the voltage ${{v}_{C}}\left( t \right)$ across the equivalent capacitance of the PEA over time when the drive voltage $u\left( t \right)$ and the hysteresis voltage ${{v}_{h}}\left( t \right)$ remain constant. In this paper, the creep effect is equivalently modeled using a spring-damper system \cite{croft2001creep,gu2013motion}:
\begin{equation}
	{{G}_{CRP}}(s)=\frac{{{V}_{C}}(s)}{{{V}_{H}}(s)}=\prod\limits_{i}{\frac{s+{{z}_{i}}}{s+{{p}_{i}}}}
	\label{eq:5}
\end{equation}
where ${{V}_{C}}\left( s \right)$ is the Laplace transform of ${{v}_{C}}\left( t \right)$. The variable $i$ represents the order of the creep model, while ${{z}_{i}}$ and ${{p}_{i}}$ are the zeros and poles of the creep transfer function, respectively.

\subsection{PFSM Electromechanical Component} \label{sec:3.2}
The electromechanical components of the PFSM, denoted as ${{G}_{EM}}$, can be divided into two parts: the inverse piezoelectric effect ${{G}_{IPE}}$ of the PEA, followed by the mechanical dynamics ${{G}_{MECH}}$, involving the PEA, flexible hinges, mounting platform, and mirror.

\subsubsection{Inverse Piezoelectric Effect} \label{sec:3.2.1}
The charge $Q$ on the equivalent capacitance of the PEA is converted into force $F$ through a linear electromechanical transformation. The inverse piezoelectric effect ${{G}_{IPE}}$ is expressed as follows:
\begin{equation}
	{{G}_{IPE}}(s)=\frac{F(s)}{{{V}_{C}}(s)}=\frac{Q(s)}{{{V}_{C}}(s)}\frac{F(s)}{Q(s)}={{C}_{PEA}}{{k}_{F}}
	\label{eq:6}
\end{equation}
where ${{C}_{PEA}}$ and ${{k}_{F}}$ represent the equivalent capacitance and the electromechanical conversion coefficient of the PEA, respectively.

\subsubsection{Mechanical Component} \label{sec:3.2.2}
\begin{figure}[htbp!]
	\centering\includegraphics[width=0.7\columnwidth]{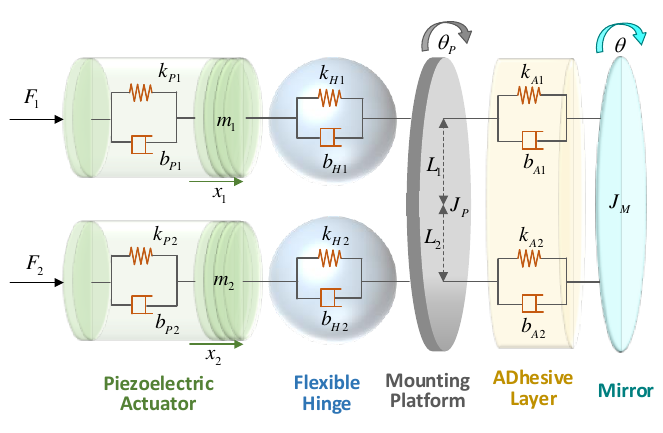}
	\caption{The mechanical model of the PFSM, represented by spring-damper dynamics.}
	\label{fig:3}
\end{figure}

The mechanical model of the PFSM is illustrated in Fig. \ref{fig:3}. The spring-damper dynamics represent the relationships among the PEA, flexible hinge, mounting platform, and mirror.

The forces ${{F}_{1}}$ and ${{F}_{2}}$ represent the output forces of PEA1 and PEA2, respectively. The equivalent masses of PEA1 and PEA2 are denoted as ${{m}_{1}}$ and ${{m}_{2}}$. The equivalent stiffness and damping of PEA1 and PEA2 are represented by ${{k}_{P1}}$,${{b}_{P1}}$ and ${{k}_{P2}}$,${{b}_{P2}}$, respectively. The equivalent stiffness and damping of the flexible hinges are denoted as ${{k}_{H1}}$,${{b}_{H1}}$ and ${{k}_{H2}}$,${{b}_{H2}}$, respectively. The equivalent stiffness and damping of the adhesive between the mounting platform and the mirror are ${{k}_{A1}}$,${{b}_{A1}}$ and ${{k}_{A2}}$,${{b}_{A2}}$, respectively. The displacements of PEA1 and PEA2 are represented by ${{x}_{1}}$ and ${{x}_{2}}$. The lever arms driven by PEA1 and PEA2 are ${{L}_{1}}$ and ${{L}_{2}}$, respectively. The moments of inertia of the platform and the mirror are denoted as ${{J}_{P}}$ and ${{J}_{M}}$. The deflection angles of the platform and the mirror are ${{\theta }_{P}}$ and $\theta $, respectively. The mechanical dynamics are then expressed as follows:
\begin{equation}
	\left\{ \begin{aligned}
		& {{m}_{1}}{{{\ddot{x}}}_{1}}+\left( {{b}_{P1}}+{{b}_{H1}} \right){{{\dot{x}}}_{1}}+\left( {{k}_{P1}}+{{k}_{H1}} \right){{x}_{1}}-{{b}_{H1}}{{L}_{1}}{{{\dot{\theta }}}_{P}}-{{k}_{H1}}{{L}_{1}}{{\theta }_{P}}={{F}_{1}} \\ 
		& {{m}_{2}}{{{\ddot{x}}}_{2}}+\left( {{b}_{P2}}+{{b}_{H2}} \right){{{\dot{x}}}_{2}}+\left( {{k}_{P2}}+{{k}_{H2}} \right){{x}_{2}}+{{b}_{H2}}{{L}_{2}}{{{\dot{\theta }}}_{P}}+{{k}_{H2}}{{L}_{2}}{{\theta }_{P}}={{F}_{2}} \\ 
		& {{J}_{P}}{{{\ddot{\theta }}}_{P}}+\left[ \left( {{b}_{H1}}+{{b}_{A1}} \right)L_{1}^{2}+\left( {{b}_{H2}}+{{b}_{A2}} \right)L_{2}^{2} \right]{{{\dot{\theta }}}_{P}}+\left[ \left( {{k}_{H1}}+{{k}_{A1}} \right)L_{1}^{2}+\left( {{k}_{H2}}+{{k}_{A2}} \right)L_{2}^{2} \right]{{\theta }_{P}} \\ 
		& \ \ \ \ \ \ \ =\left( {{b}_{A1}}L_{1}^{2}+{{b}_{A2}}L_{2}^{2} \right)\dot{\theta }+\left( {{k}_{A1}}L_{1}^{2}+{{k}_{A2}}L_{2}^{2} \right)\theta +{{b}_{H1}}{{L}_{1}}{{{\dot{x}}}_{1}}+{{k}_{H1}}{{L}_{1}}{{x}_{1}}-{{b}_{H2}}{{L}_{2}}{{{\dot{x}}}_{2}}-{{k}_{H2}}{{L}_{2}}{{x}_{2}} \\ 
		& {{J}_{M}}\ddot{\theta }+\left( {{b}_{A1}}L_{1}^{2}+{{b}_{A2}}L_{2}^{2} \right)\left( \dot{\theta }-{{{\dot{\theta }}}_{P}} \right)+\left( {{k}_{A1}}L_{1}^{2}+{{k}_{A2}}L_{2}^{2} \right)\left( \theta -{{\theta }_{P}} \right)=0 \\ 
	\end{aligned} \right.
	\label{eq:7}
\end{equation}

Let the input be $F={{\left[ {{F}_{1}},{{F}_{2}} \right]}^{T}}$, and the state be $Z={{[{{Z}_{1}},{{Z}_{2}},{{Z}_{3}},{{Z}_{4}},{{Z}_{5}},{{Z}_{6}},{{Z}_{7}},{{Z}_{8}}]}^{T}}={{[{{x}_{1}},{{\dot{x}}_{1}},{{x}_{2}},{{\dot{x}}_{2}},{{\theta }_{P}},{{\dot{\theta }}_{P}},\theta ,\dot{\theta }]}^{T}}$. The state-space representation of the system state equations and output equations can be described using vector matrices as follows:
\begin{equation}
	\left\{ \begin{aligned}
		& \dot{\mathbf{Z}}=\mathbf{AZ}+\mathbf{BF} \\ 
		& \theta =\mathbf{CZ} \\ 
	\end{aligned} \right.
	\label{eq:8}
\end{equation}
where 

$\mathbf{A}=\left[ \begin{matrix}
	{{\mathbf{A}}_{1}} & {{\mathbf{A}}_{2}}  \\
	{{\mathbf{A}}_{3}} & {{\mathbf{A}}_{4}}  \\
\end{matrix} \right]$,
${{\mathbf{A}}_{1}}=\left[ \begin{matrix}
	0 & 1 & 0 & 0  \\
	-\frac{{{k}_{P1}}+{{k}_{H1}}}{{{m}_{1}}} & -\frac{{{b}_{P1}}+{{b}_{H1}}}{{{m}_{1}}} & 0 & 0  \\
	0 & 0 & 0 & 1  \\
	0 & 0 & -\frac{{{k}_{P2}}+{{k}_{H2}}}{{{m}_{2}}} & -\frac{{{b}_{P2}}+{{b}_{H2}}}{{{m}_{2}}}  \\
\end{matrix} \right]$,

${{\mathbf{A}}_{2}}=\left[ \begin{matrix}
	0 & 0 & 0 & 0  \\
	\frac{{{k}_{H1}}{{L}_{1}}}{{{m}_{1}}} & \frac{{{b}_{H1}}{{L}_{1}}}{{{m}_{1}}} & 0 & 0  \\
	0 & 0 & 0 & 0  \\
	-\frac{{{k}_{H2}}{{L}_{2}}}{{{m}_{2}}} & -\frac{{{b}_{H2}}{{L}_{2}}}{{{m}_{2}}} & 0 & 0  \\
\end{matrix} \right]$,
${{\mathbf{A}}_{3}}=\left[ \begin{matrix}
	0 & 0 & 0 & 0  \\
	\frac{{{k}_{H1}}{{L}_{1}}}{{{J}_{P}}} & \frac{{{b}_{H1}}{{L}_{1}}}{{{J}_{P}}} & -\frac{{{k}_{H2}}{{L}_{2}}}{{{J}_{P}}} & -\frac{{{b}_{H2}}{{L}_{2}}}{{{J}_{P}}}  \\
	0 & 0 & 0 & 0  \\
	0 & 0 & 0 & 0  \\
\end{matrix} \right]$,

${{\mathbf{A}}_{4}}=\left[ \begin{matrix}
	0 & 1 & 0 & 0  \\
	-\frac{\left( {{k}_{H1}}+{{k}_{A1}} \right)L_{1}^{2}+\left( {{k}_{H2}}+{{k}_{A2}} \right)L_{2}^{2}}{{{J}_{P}}} & -\frac{\left( {{b}_{H1}}+{{b}_{A1}} \right)L_{1}^{2}+\left( {{b}_{H2}}+{{b}_{A2}} \right)L_{2}^{2}}{{{J}_{P}}} & \frac{{{k}_{A1}}L_{1}^{2}+{{k}_{A2}}L_{2}^{2}}{{{J}_{P}}} & \frac{{{b}_{A1}}L_{1}^{2}+{{b}_{A2}}L_{2}^{2}}{{{J}_{P}}}  \\
	0 & 0 & 0 & 1  \\
	\frac{{{k}_{A1}}L_{1}^{2}+{{k}_{A2}}L_{2}^{2}}{{{J}_{M}}} & \frac{{{b}_{A1}}L_{1}^{2}+{{b}_{A2}}L_{2}^{2}}{{{J}_{M}}} & -\frac{{{k}_{A1}}L_{1}^{2}+{{k}_{A2}}L_{2}^{2}}{{{J}_{M}}} & -\frac{{{b}_{A1}}L_{1}^{2}+{{b}_{A2}}L_{2}^{2}}{{{J}_{M}}}  \\
\end{matrix} \right]$,

$\mathbf{B}={{\left[ \begin{matrix}
			0 & {1}/{{{m}_{1}}} & 0 & 0 & 0 & 0 & 0 & 0  \\
			0 & 0 & 0 & {1}/{{{m}_{2}}} & 0 & 0 & 0 & 0  \\
		\end{matrix} \right]}^{T}}$,
$\mathbf{C}=\left[ \begin{matrix}
	0 & 0 & 0 & 0 & 0 & 0 & 1 & 0  \\
\end{matrix} \right]$.

This is an 8th-order multiple input single output (MISO) system with two transfer functions: one mapping from ${{F}_{1}}$ to $\theta $, and another from ${{F}_{2}}$ to $\theta $. Consequently, ${{F}_{1}}$ and ${{F}_{2}}$ are coupled, making it impossible to uniquely determine the mechanical dynamics based solely on $\theta $.

In fact, to ensure the single-axis deflection performance of the PFSM, the two selected PEAs have minimal physical structural differences. Therefore, it is reasonable to assume that the mechanical parameters related to PEA1 and PEA2 are identical, i.e., $m={{m}_{1}}={{m}_{2}}$, ${{k}_{P}}={{k}_{P1}}={{k}_{P2}}$, ${{b}_{P}}={{b}_{P1}}={{b}_{P2}}$, ${{k}_{H}}={{k}_{H1}}={{k}_{H2}}$, ${{b}_{H}}={{b}_{H1}}={{b}_{H2}}$, ${{k}_{A}}={{k}_{A1}}={{k}_{A2}}$, ${{b}_{A}}={{b}_{A1}}={{b}_{A2}}$, and $L={{L}_{1}}={{L}_{2}}$. Then let the input be $\Delta F={{F}_{1}}-{{F}_{2}}$ and the state be $Z={{[{{Z}_{1}},{{Z}_{2}},{{Z}_{3}},{{Z}_{4}},{{Z}_{5}},{{Z}_{6}}]}^{T}}={{[{{x}_{1}}-{{x}_{2}},{{\dot{x}}_{1}}-{{\dot{x}}_{2}},{{\theta }_{P}},{{\dot{\theta }}_{P}},\theta ,\dot{\theta }]}^{T}}$, the simplified mechanical dynamics equations are represented as:
\begin{equation}
	\left\{ \begin{aligned}
		& m\left( {{{\ddot{x}}}_{1}}-{{{\ddot{x}}}_{2}} \right)+\left( {{b}_{P}}+{{b}_{H}} \right)\left( {{{\dot{x}}}_{1}}-{{{\dot{x}}}_{2}} \right)+\left( {{k}_{P}}+{{k}_{H}} \right)\left( {{x}_{1}}-{{x}_{2}} \right)=2{{b}_{H}}L{{{\dot{\theta }}}_{P}}+2{{k}_{H}}L{{\theta }_{P}}+\Delta F \\ 
		& {{J}_{P}}{{{\ddot{\theta }}}_{P}}+2\left( {{b}_{H}}+{{b}_{A}} \right){{L}^{2}}{{{\dot{\theta }}}_{P}}+2\left( {{k}_{H}}+{{k}_{A}} \right){{L}^{2}}{{\theta }_{P}}=2{{b}_{A}}{{L}^{2}}\dot{\theta }+2{{k}_{A}}{{L}^{2}}\theta +{{b}_{H}}L\left( {{{\dot{x}}}_{1}}-{{{\dot{x}}}_{2}} \right)+{{k}_{H}}L\left( {{x}_{1}}-{{x}_{2}} \right) \\ 
		& {{J}_{M}}\ddot{\theta }+2{{b}_{A}}{{L}^{2}}\dot{\theta }+2{{k}_{A}}{{L}^{2}}\theta =2{{b}_{A}}{{L}^{2}}{{{\dot{\theta }}}_{P}}+2{{k}_{A}}{{L}^{2}}{{\theta }_{P}} \\ 
	\end{aligned} \right.
	\label{eq:9}
\end{equation}

It can be observed that the 8th-order MISO system (\ref{eq:8}) is simplified to a 6th-order SISO system, effectively decoupling ${{F}_{1}}$ and ${{F}_{2}}$, as well as allowing the mechanical dynamics to be uniquely identifiable. The corresponding state-space vector-matrix form is

$\mathbf{A}=\left[ \begin{matrix}
	0 & 1 & 0 & 0 & 0 & 0  \\
	-\frac{{{k}_{P}}+{{k}_{H}}}{m} & -\frac{{{b}_{P}}+{{b}_{H}}}{m} & \frac{2{{k}_{H}}L}{m} & \frac{2{{b}_{H}}L}{m} & 0 & 0  \\
	0 & 0 & 0 & 1 & 0 & 0  \\
	\frac{{{k}_{H}}L}{{{J}_{P}}} & \frac{{{b}_{H}}L}{{{J}_{P}}} & -\frac{2\left( {{k}_{H}}+{{k}_{A}} \right){{L}^{2}}}{{{J}_{P}}} & -\frac{2\left( {{b}_{H}}+{{b}_{A}} \right){{L}^{2}}}{{{J}_{P}}} & \frac{2{{k}_{A}}{{L}^{2}}}{{{J}_{P}}} & \frac{2{{b}_{A}}{{L}^{2}}}{{{J}_{P}}}  \\
	0 & 0 & 0 & 0 & 0 & 1  \\
	0 & 0 & \frac{2{{k}_{A}}{{L}^{2}}}{{{J}_{M}}} & \frac{2{{b}_{A}}{{L}^{2}}}{{{J}_{M}}} & -\frac{2{{k}_{A}}{{L}^{2}}}{{{J}_{M}}} & -\frac{2{{b}_{A}}{{L}^{2}}}{{{J}_{M}}}  \\
\end{matrix} \right]$,

$\mathbf{B}={{\left[ \begin{matrix}
			0 & {1}/{m}\; & 0 & 0 & 0 & 0  \\
		\end{matrix} \right]}^{T}}$,
$\mathbf{C}=\left[ \begin{matrix}
	0 & 0 & 0 & 0 & 1 & 0  \\
\end{matrix} \right]$.

Assuming that the inverse piezoelectric parameters of PEA1 and PEA2 are identical, i.e., ${{k}_{F1}}={{k}_{F2}}$ and ${{C}_{PEA1}}={{C}_{PEA2}}$, combining with Eq. (\ref{eq:6}), we get $\Delta F={{F}_{1}}-{{F}_{2}}=\left( {{V}_{C1}}-{{V}_{C2}} \right){{G}_{IPE}}=\Delta {{V}_{C}}{{G}_{IPE}}$, thereby obtaining the electromechanical coupled dynamics of the PFSM as follows:
\begin{equation}
	{{G}_{MECH}}(s)=\frac{\theta (s)}{\Delta F(\text{s})}=\frac{{{b}_{3}}{{s}^{3}}+{{b}_{2}}{{s}^{2}}+{{b}_{1}}s+{{b}_{0}}}{{{a}_{6}}{{s}^{6}}+{{a}_{5}}{{s}^{5}}+{{a}_{4}}{{s}^{4}}+{{a}_{3}}{{s}^{3}}+{{a}_{2}}{{s}^{2}}+{{a}_{1}}s+{{a}_{0}}}
	\label{eq:10}
\end{equation}

\subsection{Dual-axis System Modeling} \label{sec:3.3}
Creep is an inherent characteristic of piezoelectric materials, it is reasonable to assume that the creep dynamics of PEA1 and PEA2 are identical. Thus, the transfer function of the single-axis SISO system is derived as follows:
\begin{equation}
	{{G}_{SISO}}=\frac{\theta (s)}{U(s)}=\frac{\Delta {{V}_{H}}(s)}{U(s)}\frac{\Delta {{V}_{C}}(s)}{\Delta {{V}_{H}}(s)}\frac{\theta (s)}{\Delta {{V}_{C}}(s)}=\Delta {{G}_{HYS}}{{G}_{CRP}}{{G}_{EM}}
	\label{eq:11}
\end{equation}
where $\Delta {{V}_{H}}(s)={{V}_{H1}}(s)-{{V}_{H2}}(s)$, ${{V}_{H1}}(s)$ and ${{V}_{H2}}(s)$ represent the Laplace transforms of the hysteresis output voltages ${{v}_{h1}}\left( t \right)$ and ${{v}_{h2}}\left( t \right)$ of PEA1 and PEA2, respectively. $\Delta {{G}_{HYS}}={{G}_{HYS1}}-{{G}_{HYS2}}$ represents the transfer function from the drive voltage $u\left( t \right)$ to the difference in hysteresis voltages of PEA1 and PEA2, where ${{G}_{HYS1}}(s)={{{V}_{H1}}(s)}/{U(s)}$ and ${{G}_{HYS2}}(s)={{{V}_{H2}}(s)}/{U(s)}$.
\begin{figure}[htbp!]
	\centering\includegraphics[width=0.8\columnwidth]{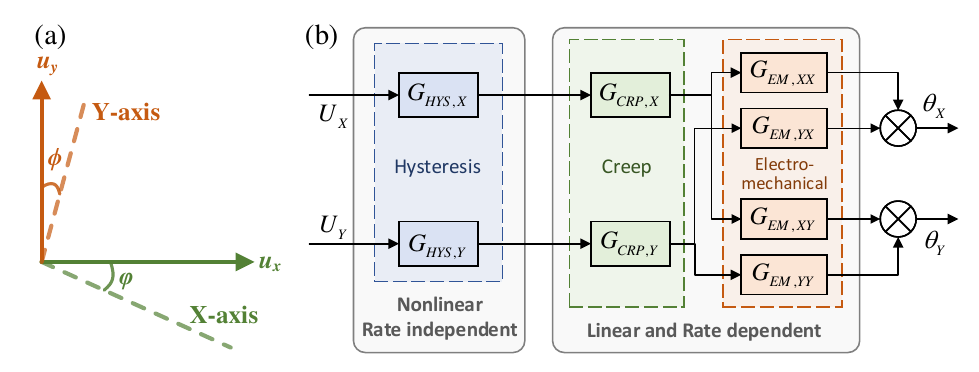}
	\caption{Diagram of the dual-axis PFSM system. (a) Geometric illustration of the dual -axis cross-coupling. (b) Block of the MIMO model.}
	\label{fig:4}
\end{figure}

The deflection angle of the PFSM exhibits two degrees of freedom. Due to the interconnection of the mechanical components, there is a correlation between the outputs of the dual axes. Fig. \ref{fig:4}(a) illustrates the cross influence. The voltages ${{u}_{x}}(t)$ and ${{u}_{y}}(t)$, driving the X and Y axis deflections, are orthogonal to each other. However, due to assembly errors and other factors, the actual X-axis direction does not perfectly align with ${{u}_{x}}$, and the same applies to the Y-axis. This misalignment introduces an additional $\cos \varphi $ gain factor in the transfer function (\ref{eq:11}) and can be incorporated into ${{G}_{EM}}$. Additionally, the X-axis and Y-axis are not strictly perpendicular, meaning deflection in one axis can induce a slight response in another axis. If assume ${{\theta }_{x}}$ and ${{\theta }_{y}}$ are linearly related in the time domain \cite{han2022design}, the frequency response of cross-axis is identical to that of the single-axis, reducing the precision of the cross-coupling modeling.

To characterize the nonlinear correlation between ${{\theta }_{x}}$ and ${{\theta }_{y}}$ under single-axis excitation, we incorporate cross-axis transfer functions ${{G}_{EM,XY}}$ (X to Y) and ${{G}_{EM,YX}}$ (Y to X), without exceeding 6th-order. This enables comprehensive modeling of the dual-axis mechanical cross-coupling. The schematic of the dual-axis MIMO system is depicted in Fig. \ref{fig:4}(b), with the mathematical model given by:
\begin{equation}
	\left[ \begin{matrix}
		{{\theta }_{X}}(s)  \\
		{{\theta }_{Y}}(s)  \\
	\end{matrix} \right]=\left[ \begin{matrix}
		\Delta {{G}_{PEA,X}}{{G}_{EM,XX}} & \Delta {{G}_{PEA,Y}}{{G}_{EM,YX}}  \\
		\Delta {{G}_{PEA,X}}{{G}_{EM,XY}} & \Delta {{G}_{PEA,Y}}{{G}_{EM,YY}}  \\
	\end{matrix} \right]\left[ \begin{matrix}
		{{U}_{X}}(s)  \\
		{{U}_{Y}}(s)  \\
	\end{matrix} \right]
	\label{eq:12}
\end{equation}
where ${{G}_{MIMO}}$ represents $2\times 2$ transfer function matrix on the right-hand side. $\Delta {{G}_{PEA,X}}=\Delta {{G}_{HYS,X}}{{G}_{CRP,X}}$ and $\Delta {{G}_{PEA,Y}}=\Delta {{G}_{HYS,Y}}{{G}_{CRP,Y}}$ correspond to the differences in voltage between the equivalent capacitors of PEA1 and PEA2 for the X and Y axes, respectively.

\section{Experiment} \label{sec:4}
\begin{figure}[htbp!]
	\centering\includegraphics[width=0.7\columnwidth]{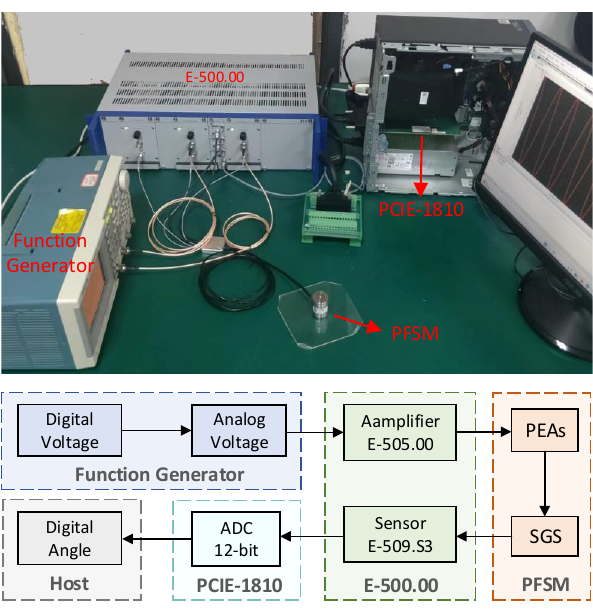}
	\caption{The block diagram of the experimental platform setup.}
	\label{fig:5}
\end{figure}

An experimental platform for the PFSM is constructed, as illustrated in Fig. \ref{fig:5}. The platform comprises a modular piezoelectric controller (E-500.00, PI Company), a dual-axis PFSM with SGS (S-330.2SL, PI Company), a 2GS/s function generator (AFG3252, Tektronix Company), a 12-bit resolution A/D data acquisition card (PCIE-1810, AdvanTech Company), and a host computer. During the experiment, the function generator produces digital signals that are converted to 0-10V analog voltages through DAC and then amplified to 0-100V by the amplifier module E-505.00. The SGS detects the structural deformation caused by the PFSM deflection, with its resistance change being converted into an analog voltage by the sensor module E-509.S3. This analog signal is then digitized to a 0-10V signal by the ADC of the PCIE-1810, and the program running on the computer linearizes it to a deflection angle range of 2mrad.

\subsection{System Identification} \label{sec:4.1}
The composite modeling of the PFSM system is developed through the study of the complex dynamic behavior of various physical components. The next step is to identify the hysteresis and creep characteristics of the PEA, along with the electromechanical dynamics of the PFSM. Fig. 6 qualitatively illustrates the dynamics responses of each module within different frequency ranges. Bouc-Wen hysteresis is a nonlinear effect with global memory, which is invariant across the entire frequency range but influenced by changes in amplitude. The creep effect of piezoelectric materials manifests slowly, affecting only the low-frequency range and is not influenced by amplitude variations. The electromechanical dynamics exhibit high bandwidth, which remain consistent within the low to mid-frequency ranges but attenuate in the high-frequency range, with no dependence on amplitude changes.
\begin{figure}[htbp!]
	\centering\includegraphics[width=0.5\columnwidth]{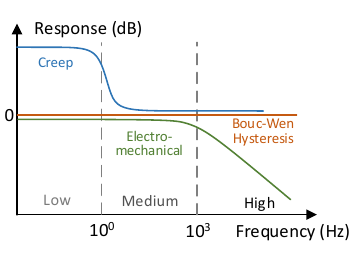}
	\caption{The dynamic responses of each module across different frequency ranges.}
	\label{fig:6}
\end{figure}

The accuracy of the identification model is evaluated based on the RMSE (Root Mean Square Relative Errors), defined as follows:
\begin{equation}
	RMSE=\sqrt{{\sum\limits_{i=1}^{N}{{{\left( {{{\hat{\theta }}}_{i}}-{{\theta }_{i}} \right)}^{2}}}}/{\sum\limits_{i=1}^{N}{\theta _{i}^{2}}}}
	\label{eq:13}
\end{equation}
where $N$ is the number of samples, $\theta $ and $\hat{\theta }$ represent the experimental measurement angle and the model prediction angle, respectively.

\subsubsection{Hysteresis Identification} \label{sec:4.1.1}
To identify the hysteresis parameters of the PEA1 and PEA2, it is essential to determine its hysteresis output voltages ${{v}_{h1}}$ and ${{v}_{h2}}$. However, since the PEA is an internal component of the PFSM, the hysteresis output voltages cannot be directly measured. The only observable variable is the deflection angle $\theta $. Under mid-frequency signal excitation, the mechanical dynamics exhibit no change in amplitude and no phase delay, making the input $\Delta {{v}_{C}}$ proportional to the output $\theta $. Additionally, the frequency response of the creep dynamic is 0 dB with no phase delay, thus $\Delta {{v}_{h}}\propto \Delta {{v}_{C}}\propto \theta $. However, the proportional coefficient is unknown, making it impossible to obtain accurate $\Delta {{v}_{h}}$. Meanwhile, there are infinite combinations of ${{v}_{h1}}$ and ${{v}_{h2}}$ that result the same $\Delta {{v}_{h}}={{v}_{h1}}-{{v}_{h2}}$, leading to the same $\theta $ under the proposed model. Therefore, ${{v}_{h1}}$ and ${{v}_{h2}}$ cannot be uniquely determined.

The simplest and most intuitive approach is to assume ${{v}_{h1}}$ and ${{v}_{h2}}$ are linearly related to $\Delta {{v}_{h}}$. This simplification implies that under mid-frequency signal excitation, the complex nonlinear relationship between ${{v}_{h1}}$ and ${{v}_{h2}}$ and $\theta $ is reduced to a linear one. Furthermore, the Bouc-Wen model parameters uniquely correspond to the normalized trajectory shape of their output under a specified input. Therefore, under this simplified linear relationship, we can solve the problem of identifying hysteresis parameters based on the observed $\theta $. In contrast, the Prandtl-Ishlinskii model requires accurate outputs to identify parameters, which are coupled with the mechanical dynamic identification. This analysis demonstrates that utilizing the Bouc-Wen model for characterizing hysteresis is reasonable.

\begin{figure}[htbp!]
	\centering\includegraphics[width=1.0\columnwidth]{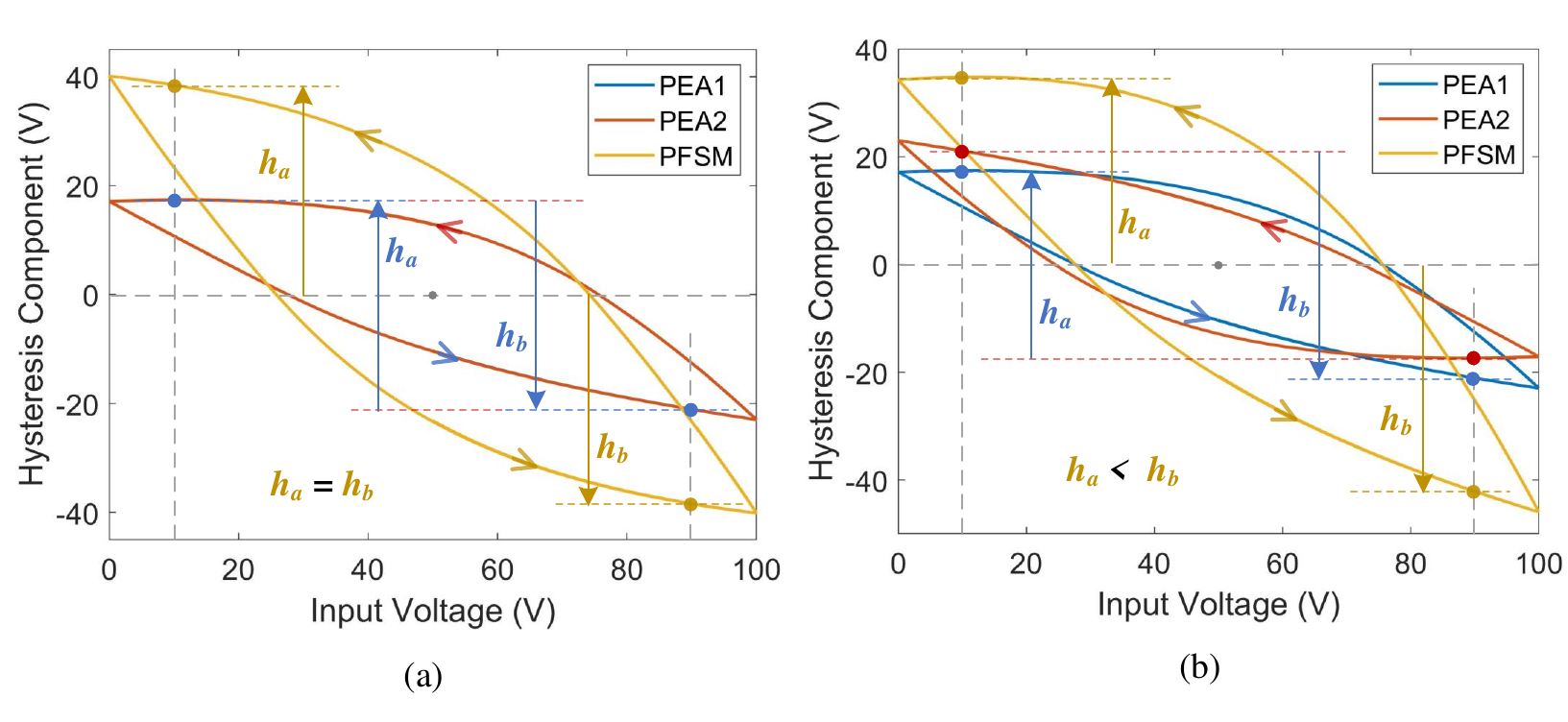}
	\caption{The hysteresis component curves of PEA1, PEA2, and single-axis PFSM. ${{h}_{a}}$ and ${{h}_{b}}$ represent the hysteresis components of the PFSM at driving voltages of 10V and 90V, respectively. (a) The hysteresis parameters of PEA1 and PEA2 are consistent, their hysteresis component curves coincide. (b) The hysteresis parameters of PEA1 and PEA2 are inconsistent, their hysteresis component curves exhibit central symmetry.}
	\label{fig:7}
\end{figure}

The physical structures of the both PEAs cannot be completely identical, resulting in some asymmetry in the hysteresis components of the PFSM. If the identified parameters from the normalized $\theta $ are uniformly applied as hysteresis parameters for both PEAs, the hysteresis component curves of PEA1 and PEA2 would coincide, as shown in Fig. \ref{fig:7}(a). For an input of 10V to PEA1, its hysteresis component is ${{h}_{1}}=18.7\text{V}$, and PEA2 has a complementary input of 90V, its hysteresis component is ${{h}_{2}}=-21.3\text{V}$, thus the hysteresis component of the PFSM is ${{h}_{a}}={{h}_{1}}-{{h}_{2}}=40\text{V}$. When an input of 10V for PEA1, the corresponding PFSM hysteresis component is ${{h}_{b}}={{h}_{1}}-{{h}_{2}}=-40\text{V}$. Since ${{h}_{a}}+{{h}_{b}}=0$, the hysteresis curve of the PFSM would be symmetrical, which contradicts the observed hysteresis asymmetry of the PFSM.

Therefore, the hysteresis parameters of PEA1 and PEA2 must be inconsistent to produce asymmetry in the hysteresis component of PFSM. Nevertheless, the values of the hysteresis components for PEA1 and PEA2 are not the primary concern, their difference $\Delta {{v}_{h}}$ is crucial as it is the sole input for the mechanical dynamics. Naturally, we can use $\theta $ to identify the hysteresis parameters of PEA1 and $-\theta $ to identify the hysteresis parameters of PEA2. This results the hysteresis component curve of PEA2 being centrally symmetric to that of PEA1 around $\left( 50,0 \right)$, as plotted in Fig. \ref{fig:7}(b). Consequently, there is ${{v}_{h1}}+{{v}_{h2}}=100$ and ${{h}_{2}}=-{{h}_{1}}$. When PEA1 is subjected to an input of 10V, its hysteresis component is ${{h}_{1}}=18.7\text{V}$, meanwhile a complementary input of 90V to PEA2 corresponds to ${{h}_{2}}=-18.7\text{V}$. Thus, the hysteresis component of the PFSM is ${{h}_{a}}={{h}_{1}}-{{h}_{2}}=37.4\text{V}$. When PEA1 is given an input of 90V, the hysteresis component of the PFSM is ${{h}_{b}}={{h}_{1}}-{{h}_{2}}=-42.6\text{V}$. Given that ${{h}_{a}}+{{h}_{b}}\ne 0$, this constructs an asymmetrical hysteresis curve for the PFSM.

Once the hysteresis parameters of PEA1 ${{\mathbf{\psi }}_{1}}=\left[ {{\alpha }_{1}},{{\beta }_{1}},{{\gamma }_{1}},{{\delta }_{1}},{{n}_{1}} \right]$ are identified, the hysteresis parameters of PEA2 ${{\mathbf{\psi }}_{2}}=\left[ {{\alpha }_{2}},{{\beta }_{2}},{{\gamma }_{2}},{{\delta }_{2}},{{n}_{2}} \right]$ can be obtained by substituting ${{u}_{2}}\left( t \right)=100-{{u}_{1}}\left( t \right)$ into Eq. (\ref{eq:3}):
\begin{equation}
	\begin{aligned}
		& {{{\dot{h}}}_{2}}(t)=-\left( {{\alpha }_{2}}+100{{\delta }_{2}} \right){{{\dot{u}}}_{1}}(t)-{{\beta }_{2}}\left| {{{\dot{u}}}_{1}}(t) \right|{{\left| {{h}_{2}}(t) \right|}^{{{n}_{2}}-1}}{{h}_{2}}(t) \\ 
		& \ \ \ \ \ \ \ \ \ \ +{{\gamma }_{2}}{{{\dot{u}}}_{1}}(t){{\left| {{h}_{2}}(t) \right|}^{{{n}_{2}}}}+{{\delta }_{2}}{{{\dot{u}}}_{1}}(t){{u}_{1}}(t) \\ 
	\end{aligned}
	\label{eq:14}
\end{equation}

Let ${{\alpha }_{2}}={{\alpha }_{1}}+100{{\delta }_{1}}$, ${{\beta }_{2}}={{\beta }_{1}}$, ${{\gamma }_{2}}={{\gamma }_{1}}$, ${{\delta }_{2}}=-{{\delta }_{1}}$, and ${{n}_{2}}={{n}_{1}}$, then obtain ${{\dot{h}}_{1}}(t)+{{\dot{h}}_{2}}(t)=0$, i.e., ${{h}_{2}}=-{{h}_{1}}$.

The objective of the identification is to minimize the difference between the normalized ${{v}_{h1}}$ and $\theta $. The nonlinear least squares method is employed to estimate the hysteresis parameters of PEA1, and the corresponding optimization problem is as follows:
\begin{equation}
	\underset{{{\mathbf{\psi }}_{1}}}{\mathop{\mathbf{\psi }_{1}^{*}=\arg \min }}\,\frac{1}{2}{{\left[ \bar{\mathbf{\theta }}-\bar{\mathbf{g}}\left( \mathbf{u},{{\mathbf{\psi }}_{1}} \right) \right]}^{T}}\left[ \bar{\mathbf{\theta }}-\bar{\mathbf{g}}\left( \mathbf{u},{{\mathbf{\psi }}_{1}} \right) \right]
	\label{eq:15}
\end{equation}
where $\mathbf{\theta }$ represents the PFSM deflection angle sequence, $\mathbf{g}\left( \mathbf{u},{{\mathbf{\psi }}_{1}} \right)={{\mathbf{v}}_{h1}}$ is calculated according to Eq. (\ref{eq:3}). The normalized sequences $\bar{\mathbf{\theta }}={\left( \mathbf{\theta }-{{\theta }_{\min }} \right)}/{\left( {{\theta }_{\max }}-{{\theta }_{\min }} \right)}$ and $\bar{\mathbf{g}}={\left( \mathbf{g}-{{g}_{\min }} \right)}/{\left( {{g}_{\max }}-{{g}_{\min }} \right)}$ are obtained by linearly mapping the minimum and maximum to $\left[ 0,1 \right]$.

The nonlinear hysteresis is rate-independent and amplitude-dependent, whereas creep and mechanical dynamics exhibit the opposite behavior. Therefore, the chosen excitation signal should cover a broad range of amplitudes at a fixed frequency that is neither too low to avoid the effects of creep nor too high to prevent triggering mechanical dynamic behavior, typically selected within the range of 1Hz to 20Hz. Additionally, the amplitude of the excitation signal should span 10\% to 90\% of the control range to cover a greater extent of the travel. Consequently, this work employs excitation signals with maximum amplitude of 8V, corresponding to amplifier output voltages of 10V-90V, with frequencies of 5Hz and 10Hz, i.e., ${{u}_{a}}\left( t \right)=50+40\sin \left( 10\pi t \right)\cos \left( 0.5\pi t \right)$ and ${{u}_{b}}\left( t \right)=50+40\sin \left( 20\pi t \right)\cos \left( 0.5\pi t \right)$ for $t\in \left[ 0,1s \right]$, with a sampling time step of $10\mu s$, applied to the X-axis for identification, and similarly to the Y-axis.

Based on the deflection angle $\mathbf{\theta }$, the ‘lsqcurvefit’ function from the Matlab Optimization Toolbox is utilized to identify the classical Bouc-Wen (B-W) model \cite{ismail2009hysteresis}, the asymmetric B-W model \cite{zhu2023modeling}, and the improved asymmetric B-W model proposed in this paper, as described in Eq. (\ref{eq:3}). Under the excitation signals ${{u}_{a}}$ and ${{u}_{b}}$, the identified parameters for the three models, along with the RMSE between the normalized model output $\bar{\mathbf{v}}_{h1}$ and the normalized experimental output $\bar{\mathbf{\theta }}$, are listed in Tab. \ref{tab:1}. It can be found that the classical B-W model \cite{ismail2009hysteresis} has the largest error due to its lack of asymmetry. The asymmetric B-W model \cite{zhu2023modeling} shows significant variations in identified parameters and RMSE under different frequency excitations due to its lack of frequency independence. Our proposed asymmetric B-W model exhibits the best consistency, with minimal differences in identified parameters under different frequency excitations, and the smallest RMSE between the model prediction and actual output.
\begin{table}[htpb!]
	\centering
	\caption{RMSE of different Bouc-Wen models under two excitation signals on the X-axis}
	\label{tab:1}
	\begin{tabular}{cccccccc}
		\hline
		\multirow{2}{*}{\begin{tabular}[c]{@{}c@{}}Hysteresis Model	\end{tabular}}           & \multirow{2}{*}{\begin{tabular}[c]{@{}c@{}}Excitation Signals \end{tabular}} & \multicolumn{5}{c}{Parameters}                       & \multirow{2}{*}{RMSE} \\ \cline{3-7}
		&                                                                  & $\alpha $       & $\beta $       & $\gamma $        & $\delta $        & $n$    &                       \\ \hline
		\multirow{2}{*}{\begin{tabular}[c]{@{}c@{}}Classic B-W \cite{ismail2009hysteresis}\end{tabular}}  & $u_a$                                                            & -0.4056 & 5.6E-03 & -4.2E-03 & --       & 1.69 & 0.0106                \\
		& $u_b$                                                            & -0.4091 & 5.9E-03 & -4.4E-03 & --       & 1.66 & 0.0104                \\ \hline
		\multirow{2}{*}{\begin{tabular}[c]{@{}c@{}}Asymmetric B-W \cite{zhu2023modeling}\end{tabular}} & $u_a$                                                            & -0.3683 & 8.5E-03 & -4.8E-03 & -4.2E-05 & 1.41 & 0.0063                \\
		& $u_b$                                                            & -0.3705 & 8.2E-03 & -4.1E-03 & -7.8E-05 & 1.45 & 0.0068                \\ \hline
		\multirow{2}{*}{\begin{tabular}[c]{@{}c@{}}Proposed Asymmetric B-W \end{tabular}}       & $u_a$                                                            & -0.3767 & 0.0197  & -0.0173  & -1.2E-03 & 1.26 & 0.0038                \\
		& $u_b$                                                            & -0.3764 & 0.0192  & -0.0164  & -1.3E-03 & 1.28 & 0.0038                \\ \hline
	\end{tabular}
	\begin{tablenotes}
	%	\footnotesize
	\item
		E-3 and E-5 represent scientific notation, corresponding to ${{10}^{-3}}$ and ${{10}^{-5}}$, respectively.
	\end{tablenotes}
\end{table}

\begin{figure}[htbp!]
	\centering\includegraphics[width=0.9\columnwidth]{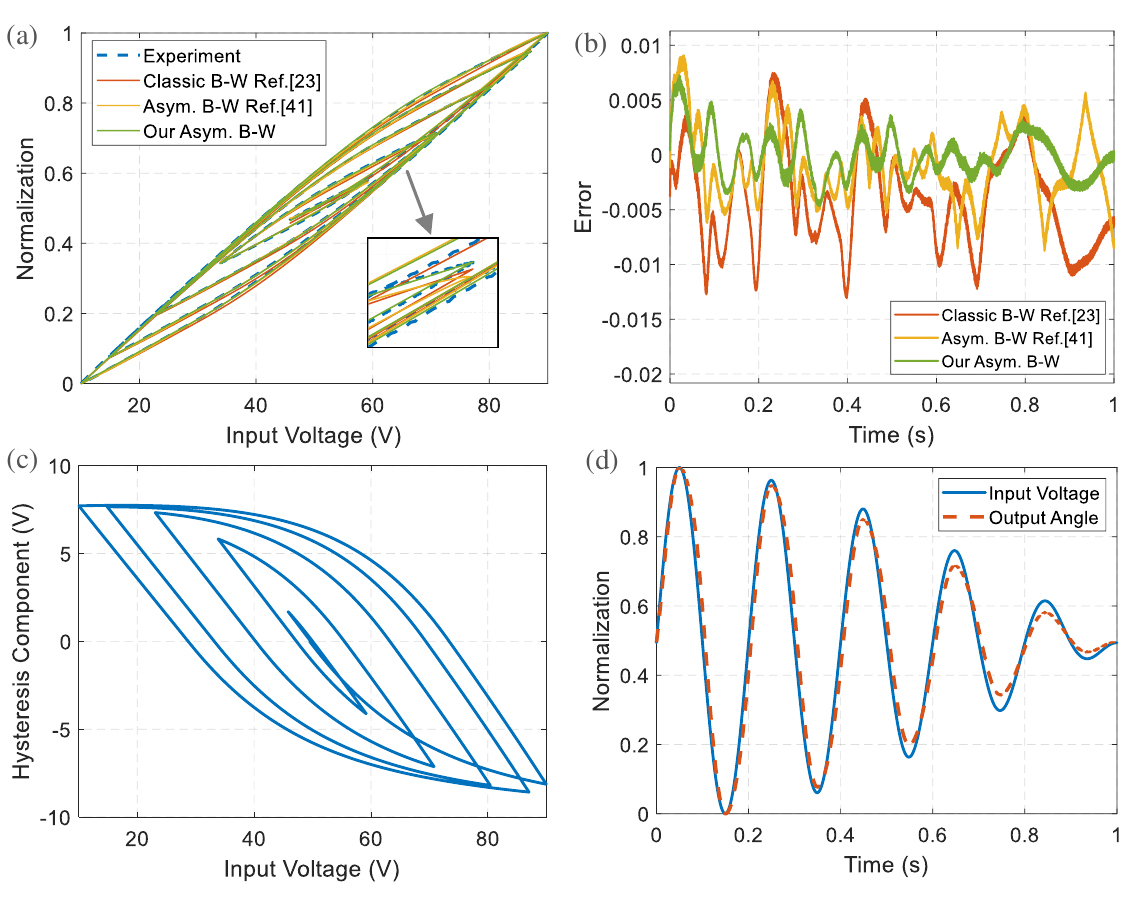}
	\caption{Performance comparison of three Bouc-Wen hysteresis models on the X-axis. (a) Normalization of predicted and experimental data as a function of input voltage. (b) Errors between predictions and experiments over time. (c) Hysteresis component of the proposed asymmetric B-W model with input voltage. (d) Normalization of input voltage and output angle over time.}
	\label{fig:8}
\end{figure}

Taking the excitation signal ${{u}_{a}}$ as an example, Fig. \ref{fig:8}(a) displays the hysteresis curves of the predicted $\bar{\mathbf{v}}_{h1}$ and the measured$\bar{\mathbf{\theta }}$ as functions of the input voltage for the three models. Fig. \ref{fig:8}(b) illustrates the errors between the predictions and measurements, presenting that the proposed asymmetric B-W model exhibits the best consistency. Additionally, Fig. \ref{fig:8}(c) plots the hysteresis curve of the proposed asymmetric B-W model, where the hysteresis components are not equal at input voltages of 10V and 90V, indicating the hysteresis asymmetry attributed to the factor $\delta $. Fig. \ref{fig:8}(d) depicts the normalization of input voltage and output angle over time, it can be observed that as the amplitude decreases, the differences between output and input at the peaks become more pronounced, demonstrating the hysteresis characteristic is amplitude-dependent.

As shown in Tab. \ref{tab:1}, the parameter identification results of the proposed asymmetric B-W model exhibit minimal differences under the two excitation signals. The RMSE with ${{u}_{a}}$ (5Hz) excitation is smaller, leading to the adoption of its results. The identified parameters for the X-axis are:
\begin{equation}
	\left\{ \begin{aligned}
		& {{\psi }_{X1}}=\left[ -0.3767,0.0197,-0.0173,-0.0012,1.16 \right] \\ 
		& {{\psi }_{X2}}=\left[ -0.4993,0.0197,-0.0173,0.0012,1.16 \right] \\ 
	\end{aligned} \right.
	\label{eq:16}
\end{equation}

For the Y-axis, the identified parameters are as follows:
\begin{equation}
	\left\{ \begin{aligned}
		& {{\psi }_{Y1}}=\left[ -0.3824,0.0209,-0.0181,-0.0012,1.13 \right] \\ 
		& {{\psi }_{Y2}}=\left[ -0.5031,0.0209,-0.0181,0.0012,1.13 \right] \\ 
	\end{aligned} \right.
	\label{eq:17}
\end{equation}

\subsubsection{Creep Identification} \label{sec:4.1.2}
The creep characteristics of the PEA are observed as a gradual change in output over time with a constant input, typically affecting the low-frequency range. Specifically, a square wave signal with a period of 80 seconds and an amplitude range of 2V to 8V (with amplifier output ranging from 20V to 80V) is adopted, given by $u\left( t \right)=50+30\ \text{square}\left( 0.025\pi t \right)$, where $t\in \left[ 0,160s \right]$, and $\text{square}\left( \cdot  \right)$ represents the 'square' function in MATLAB. Additionally, high-frequency variations at the rising and falling edges of the square wave can induce mechanical dynamics. If output data is collected during these high-frequency variations, it may reduce the accuracy of creep dynamic identification. The open-loop bandwidth of the PFSM is on the order of kHz, indicating that the step response can stabilize within 1ms. Consequently, to prevent including data from periods of rapid output fluctuations due to step excitation, the sampling time should exceed 1ms. Thus, a sampling time of 5ms is selected, which not only avoids these abrupt samples but also preserves the continuous dynamic behavior of creep.

Creep directly influences the hysteresis output voltages ${{v}_{h1}}$ and ${{v}_{h2}}$ of PEA1 and PEA2. Under the assumption that the creep characteristics of both PEAs are consistent, it follows from the SISO system transfer function (\ref{eq:11}) that the creep of the PEA is transmitted to the PFSM deflection angle. Consequently, the creep dynamics of the PEA can be indirectly identified by observing changes in the deflection angle.

\begin{figure}[htbp!]
	\centering\includegraphics[width=0.9\columnwidth]{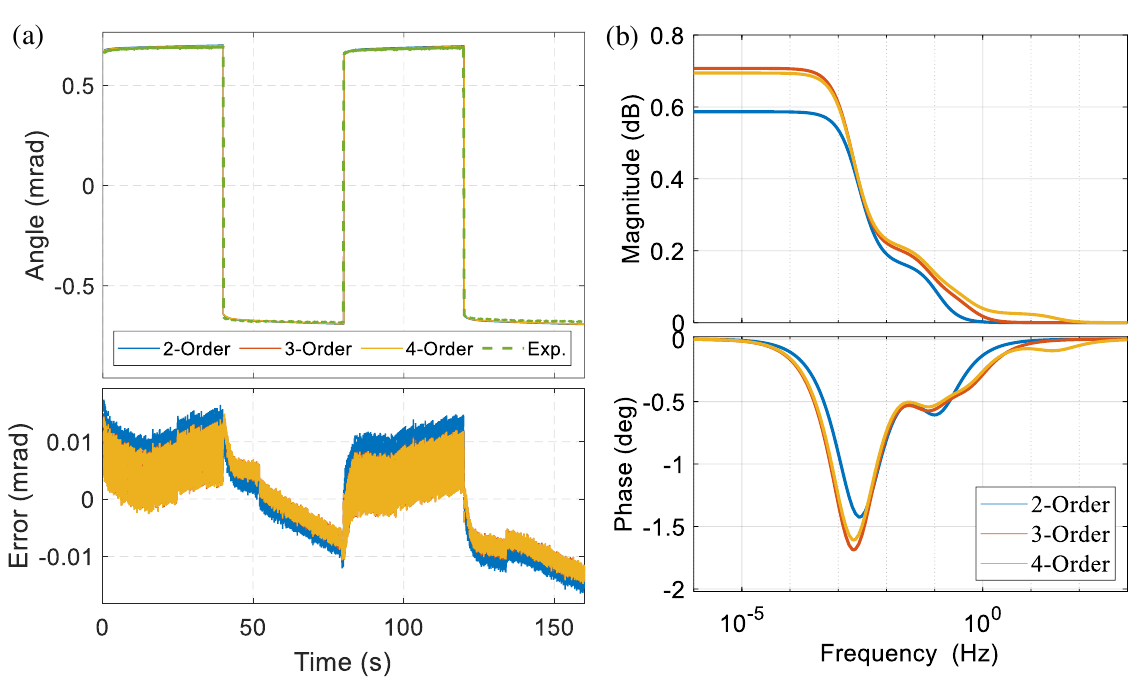}
	\caption{Performance of different order creep models for the X-axis. (a) Time-domain output and error. (b) Frequency response of the identified creep models.}
	\label{fig:9}
\end{figure}

\begin{figure}[htbp!]
	\centering\includegraphics[width=0.9\columnwidth]{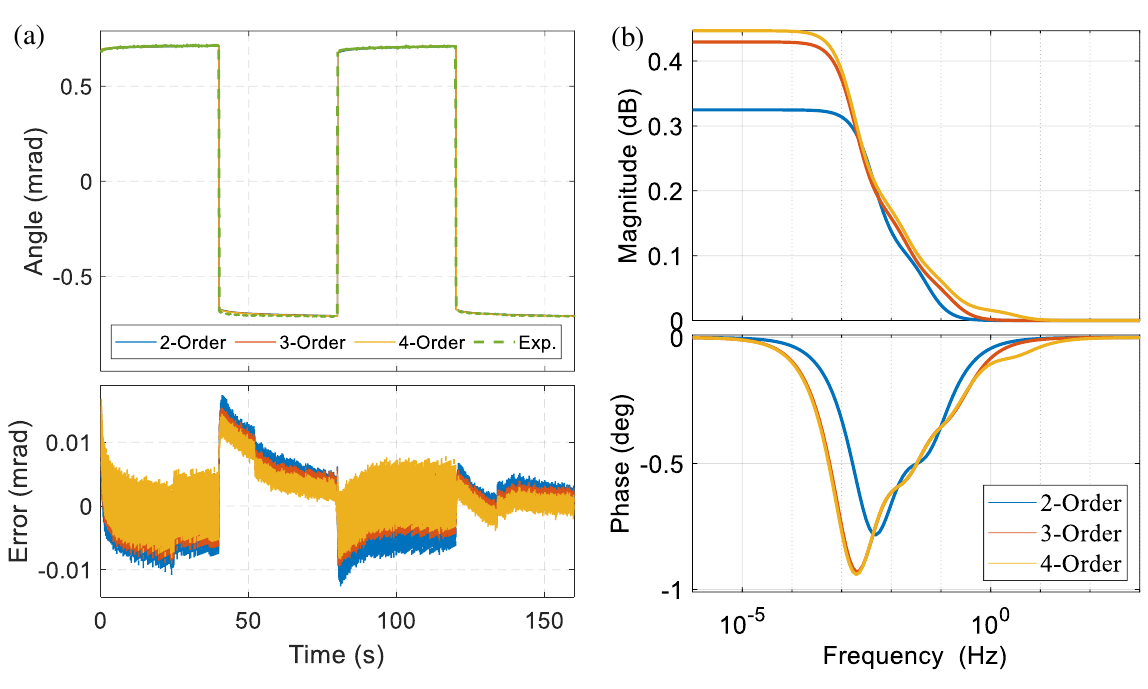}
	\caption{Performance of different order creep models for the Y-axis. (a) Time-domain output and error. (b) Frequency response of the identified creep models.}
	\label{fig:10}
\end{figure}

A square wave signal is applied separately to each axis, with the input given by$\Delta {{v}_{h}}$, calculated from the previously identified hysteresis parameters, and the output is $\mathbf{\theta }$. The ‘tfest’ function from the MATLAB Identification Toolbox is employed to identify the creep dynamics for model orders $i=2,3,4$. The identification results for the X-axis and Y-axis are presented in Figs. \ref{fig:9} and \ref{fig:10}, respectively.

Fig. \ref{fig:9}(a) compares the predicted and experimental outputs over time for different model orders, along with the prediction errors. It is observed that increasing the model order from 2 to 3 significantly reduces the error, with the RMSE for the X-axis decreasing from 0.0082 to 0.0073 and for the Y-axis decreasing from 0.0078 to 0.0065. However, the error remains relatively unchanged beyond the 3rd order, indicating that further increases in order provide minimal improvement in prediction accuracy, which is consistent with the findings reported in Ref. \cite{croft2001creep}.

Next, we eliminate the gain coefficient and convert the transfer function to a canonical form consistent with Eq. (\ref{eq:5}). Fig. \ref{fig:9}(b) presents the Bode of the identified model, showing that the creep dynamics exhibit negligible amplitude attenuation and phase delay above 10Hz. This demonstrates that creep dynamics predominantly affect the low-frequency range.

We select the identification results of the 3rd order model. The corresponding creep dynamics for the X-axis and Y-axis are as follows:
\begin{equation}
	{{G}_{CRP,X}}=\frac{{{s}^{3}}+3.787{{s}^{2}}+1.678s+0.0217}{{{s}^{3}}+3.750{{s}^{2}}+1.637s+0.0200}
	\label{eq:18}
\end{equation}

\begin{equation}
	{{G}_{CRP,Y}}=\frac{{{s}^{3}}+5.381{{s}^{2}}+4.014s+0.2482}{{{s}^{3}}+5.338{{s}^{2}}+3.933s+0.2379}
	\label{eq:19}
\end{equation}

\subsubsection{Electromechanical Identification} \label{sec:4.1.3}
Electromechanical dynamics exhibit amplitude attenuation and phase delay with increasing frequency. Therefore, the excitation signal is chosen to be a swept sine wave. To avoid the influence of amplitude-dependent hysteresis characteristics, the swept signal is set to a constant amplitude of 8V (amplifier output ranging from 10V to 90V). Additionally, the open-loop output is filtered through a 3kHz low-pass filter in the E-509 module. Hence, the sweep range is selected to cover a broad band from 1Hz to 2kHz. Consequently, the swept excitation signal is given $u=50+40\ \text{chirp}\left( t,1,2000,-{\pi }/{2}\; \right)$, for $t\in \left[ 0,60s \right]$, where $\text{chirp}\left( \cdot  \right)$ denotes the ‘chirp’ function in MATLAB. The sampling time step is set to $10\mu s$. According to Eq. (\ref{eq:12}), the electromechanical cross transfer function matrix can be extracted as follows:
\begin{equation}
	\left[ \begin{matrix}
		{{\theta }_{X}}  \\
		{{\theta }_{Y}}  \\
	\end{matrix} \right]=\left[ \begin{matrix}
		{{G}_{EM,XX}} & {{G}_{EM,YX}}  \\
		{{G}_{EM,XY}} & {{G}_{EM,YY}}  \\
	\end{matrix} \right]\left[ \begin{matrix}
		\Delta {{V}_{C,X}}  \\
		\Delta {{V}_{C,Y}}  \\
	\end{matrix} \right]
	\label{eq:20}
\end{equation}

\begin{figure}[htbp!]
	\centering\includegraphics[width=1.0\columnwidth]{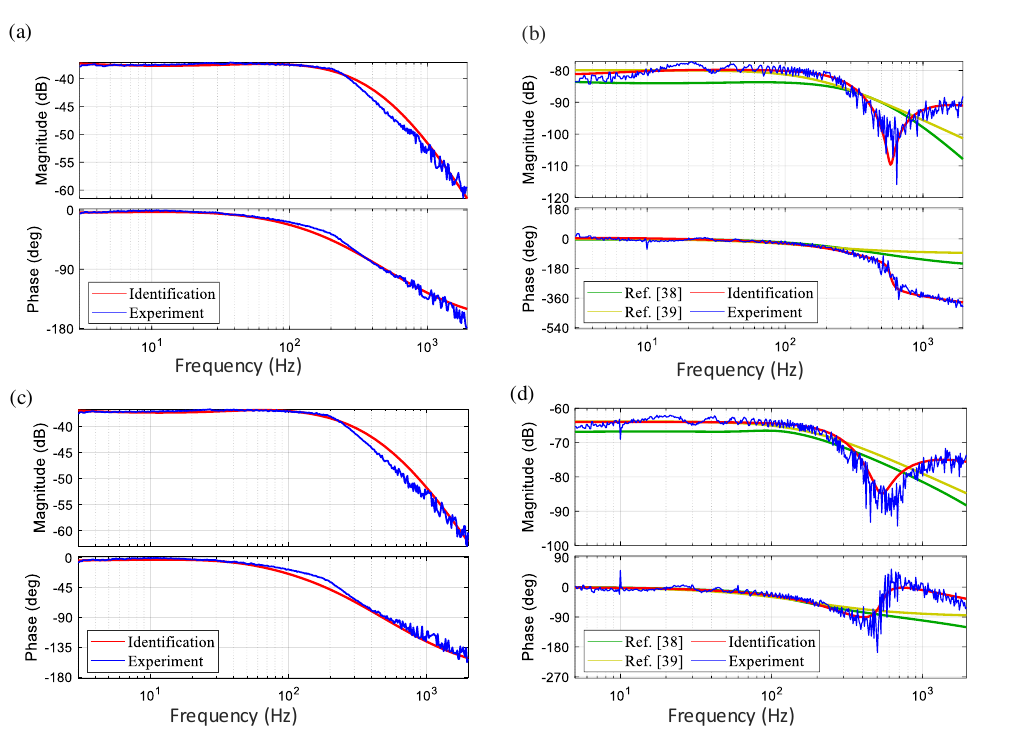}
	\caption{Frequency domain responses of the identified models and experimental outputs for mechanical dual-axis cross-coupling. (a) X-axis input to X-axis output. (b) X-axis input coupled to Y-axis output. (c) Y-axis input to Y-axis output. (d) Y-axis input coupled to X-axis output.}
	\label{fig:11}
\end{figure}

The input for identification is the voltage difference $\Delta {{V}_{C}}$ between the equivalent capacitors of PEA1 and PEA2, calculated based on the identified hysteresis and creep parameters, while the output is the deflection angle $\theta $. We use the ‘tfest’ function from the MATLAB System Identification Toolbox to estimate the transfer function based on the input-output frequency domain responses from 1Hz to 2kHz. First, the excitation signal is applied to the X-axis while the Y-axis has no control voltage. Fig. \ref{fig:11}(a) shows the frequency domain response of $\Delta {{V}_{C,X}}$ to ${{\theta }_{X}}$ and the estimated transfer function ${{G}_{EM,XX}}$. Fig. \ref{fig:11}(b) displays the residual response of $\Delta {{V}_{C,X}}$ to ${{\theta }_{Y}}$ and the estimated transfer function ${{G}_{EM,XY}}$. Next, the excitation signal is applied to the Y-axis with the X-axis voltage set to zero. Fig. \ref{fig:11}(c) shows the frequency domain response of $\Delta {{V}_{C,Y}}$ to ${{\theta }_{Y}}$ and the estimated transfer function ${{G}_{EM,YY}}$. Fig. \ref{fig:11}(d) describes the weak response of $\Delta {{V}_{C,Y}}$ to ${{\theta }_{X}}$ and the estimated transfer function ${{G}_{EM,YX}}$. Additionally, for the mechanical cross-coupling modeling, we compared our results with a linear matrix model \cite{han2022design} established in the time domain and a first-order inertia lag model \cite{wang2019comprehensive} in the frequency domain, with the corresponding frequency responses plotted in Fig. \ref{fig:11}(b) and \ref{fig:11}(d), respectively.

It can be observed that the identification models for input-output along the same axis exhibit excellent consistency, with fitting degrees reaching 89.4\% for the X-axis and 87.8\% for the Y-axis. In the time domain, the RMSE is 0.153 for the X-axis and 0.172 for the Y-axis. For cross-coupling between axes, the frequency response amplitude is relatively small and the shape is complex. Method \cite{han2022design}, which considers only linear relationships in the time domain, results in a uniformly downward-shifted amplitude-frequency response with an unchanged phase, leading to reduced consistency. Method \cite{wang2019comprehensive} employs a simpler model, which lacks precision in characterizing high-frequency features. In contrast, the proposed method achieves the best fitting consistency across the entire frequency range. Additionally, the fitting degrees in frequency domain and RMSE in time domain for the three methods are summarized in Tab. \ref{tab:2}, where the average fitting degree of the proposed method is 2.3 times that of method \cite{han2022design} and 1.3 times that of method \cite{wang2019comprehensive}. Furthermore, the proposed method reduces the RMSE by 36.2\% and 30.1\% compared to methods \cite{han2022design} and \cite{wang2019comprehensive}, respectively, confirming that the proposed method significantly improves the accuracy of cross-coupling modeling.
\begin{table}[htpb]
	\centering
	\caption{Frequency-domain fitting degree and time-domain RMSE of different methods}
	\label{tab:2}
	\begin{tabular}{@{}lccccccc@{}}
		\toprule
		\multicolumn{1}{c}{} & \multicolumn{3}{c}{Frequency-domain fitting degree (\%)}    &  & \multicolumn{3}{c}{Time-domain RMSE}        \\ \cmidrule(r){1-4} \cmidrule(l){6-8} 
		\multicolumn{1}{c}{} & \begin{tabular}[c]{@{}c@{}}Method \\ \cite{han2022design}\end{tabular}  & \begin{tabular}[c]{@{}c@{}}Method \\ \cite{wang2019comprehensive}\end{tabular} & \begin{tabular}[c]{@{}c@{}}Proposed \\ method\end{tabular}  &  & \begin{tabular}[c]{@{}c@{}}Method \\ \cite{han2022design}\end{tabular} & \begin{tabular}[c]{@{}c@{}}Method \\ \cite{wang2019comprehensive}\end{tabular} & \begin{tabular}[c]{@{}c@{}}Proposed \\ method\end{tabular}   \\ \midrule
		X to Y Axis      & 24.89       & 46.50       & 65.69 &  & 0.8494      & 0.7863      & 0.5336 \\
		Y to X Axis      & 35.48       & 54.85       & 69.84 &  & 0.7681      & 0.6998      & 0.4978 \\ \bottomrule
	\end{tabular}
\end{table}

Consequently, the estimated dual-axis cross-coupling mechanical transfer functions are as follows:
\begin{equation}
	\begin{aligned}
		{{G}_{EM,XX}}=  
		\frac{1.541\times {{10}^{11}}{{s}^{3}}+9.166\times {{10}^{13}}{{s}^{2}}+1.377\times {{10}^{16}}s+2.343\times {{10}^{17}}}{{{s}^{6}}+1.14\times {{10}^{6}}{{s}^{5}}+8.23\times {{10}^{9}}{{s}^{4}}+1.55\times {{10}^{13}}{{s}^{3}}+7.43\times {{10}^{15}}{{s}^{2}}+1.06\times {{10}^{18}}s+1.61\times {{10}^{19}}} \\ 
	\end{aligned}
	\label{eq:21}
\end{equation}
\begin{equation}
	\begin{aligned}
		{{G}_{EM,XY}}= 
		\frac{9.018\times {{10}^{4}}{{s}^{3}}-2.825\times {{10}^{7}}{{s}^{2}}+1.205\times {{10}^{12}}s+4.351\times {{10}^{13}}}{{{s}^{5}}+1.7\times {{10}^{5}}{{s}^{4}}+2.824\times {{10}^{9}}{{s}^{3}}+8.891\times {{10}^{12}}{{s}^{2}}+1.204\times {{10}^{16}}s+5.145\times {{10}^{17}}} \\ 
	\end{aligned}
	\label{eq:22}
\end{equation}
\begin{equation}
	\begin{aligned}
		{{G}_{EM,YY}}=  
		\frac{9.848\times {{10}^{10}}{{s}^{3}}+7.65\times {{10}^{13}}{{s}^{2}}+1.636\times {{10}^{16}}s+2.645\times {{10}^{17}}}{{{s}^{6}}+7.41\times {{10}^{5}}{{s}^{5}}+5.46\times {{10}^{9}}{{s}^{4}}+1.06\times {{10}^{13}}{{s}^{3}}+6.05\times {{10}^{15}}{{s}^{2}}+1.21\times {{10}^{18}}s+1.72\times {{10}^{19}}} \\ 
	\end{aligned}
	\label{eq:23}
\end{equation}
\begin{equation}
	\begin{aligned}
		{{G}_{EM,YX}}= 
		\frac{4.583{{s}^{3}}+1.301\times {{10}^{6}}{{s}^{2}}+1.145\times {{10}^{9}}s+1.447\times {{10}^{13}}}{{{s}^{4}}+4.423\times {{10}^{5}}{{s}^{3}}+6.35\times {{10}^{9}}{{s}^{2}}+1.764\times {{10}^{13}}s+2.293\times {{10}^{16}}} \\ 
	\end{aligned}
	\label{eq:24}
\end{equation}

\subsection{System Evaluation} \label{sec:4.2}
We conducted a comparative experiment on both single-axis (SISO) and dual-axis (MIMO) systems under various composite excitation signals, providing robust validation for the performance of models. The comparison is made with the existing PFSM comprehensive modeling methods \cite{liu2019composite} and \cite{wang2019comprehensive}.

\subsubsection{Single-axis SISO} \label{sec:4.2.1}
We design composite excitation signal to be applied separately to the X-axis and Y-axis while the other axis remains unexcited. The composite signal $uc(t)$ is composed of sine waves at three typical frequencies with different amplitudes, i.e., $uc\left( t \right)=50+10\sin \left( 40\pi t \right)+20\sin \left( 80\pi t \right)+15\sin \left( 240\pi t \right)$, for $t\in \left[ 0,50ms \right]$. The sampling time step is $10\mu s$.
\begin{figure}[htbp!]
	\centering\includegraphics[width=1.0\columnwidth]{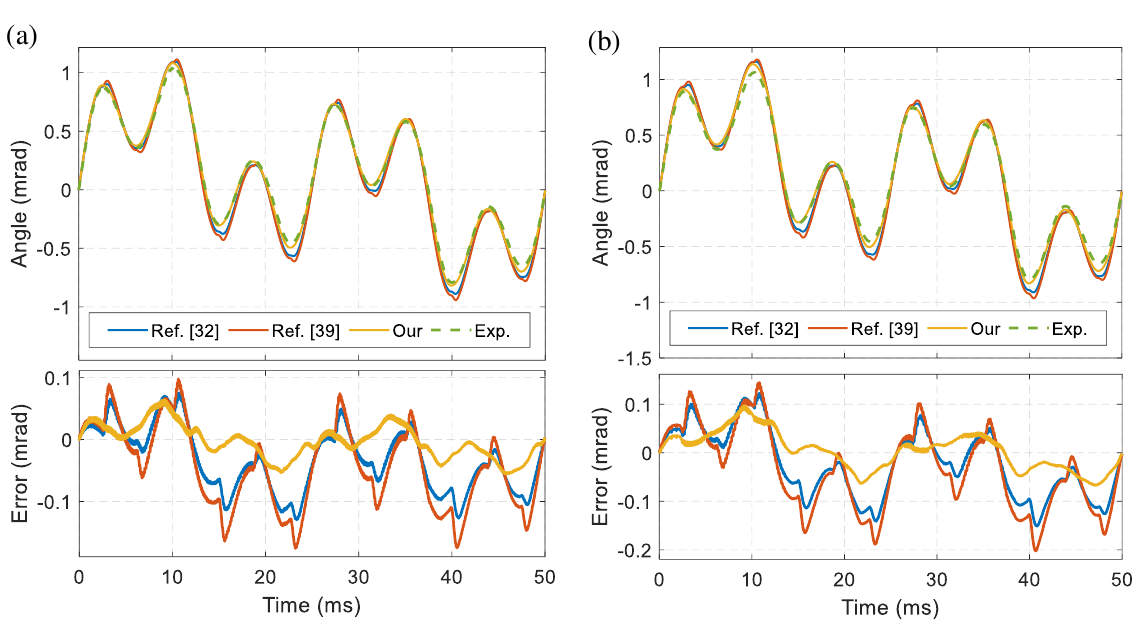}
	\caption{Output and error of the SISO model under composite signal $uc(t)$ excitation. (a) X-axis, (b) Y-axis.}
	\label{fig:12}
\end{figure}

Under $uc(t)$ excitation, the SISO model outputs for the X-axis and Y-axis, along with the experimental outputs and their errors over time, are plotted in Fig. \ref{fig:12}(a) and (b), respectively. It is evident that the output of the proposed comprehensive model aligns closely with the actual trajectory, exhibiting minimal error fluctuations. In contrast, methods \cite{liu2019composite} and \cite{wang2019comprehensive}, which both employ the classic Bouc-Wen model, fail to accurately describe the asymmetric hysteresis characteristics of PFSM, resulting in larger errors. Correspondingly, the RMSE of the outputs of the three comprehensive models for the X-axis and Y-axis are presented in Tab. 3. For the X-axis, the RMSE of the proposed method is significantly reduced by 50.1\% and 62.3\% compared to methods \cite{liu2019composite} and \cite{wang2019comprehensive}, respectively. For the Y-axis, the RMSE is reduced by 39.8\% and 48.5\% compared to methods \cite{liu2019composite} and \cite{wang2019comprehensive}, respectively. Consequently, the SISO system dynamics based on the proposed comprehensive model are satisfactory.
\begin{table}[htpb!]
	\centering
	\caption{RMSE of comprehensive SISO models under composite signal excitation}
	\label{tab:3}
	\begin{tabular}{@{}ccc@{}}
		\toprule
		Comprehensive  Modeling       & X-axis     & Y-axis     \\ \midrule
		Model \cite{liu2019composite} & 0.1215 & 0.1424 \\
		Model \cite{wang2019comprehensive} & 0.1581 & 0.1664 \\
		Proposed Model        & 0.0596 & 0.0857 \\ \bottomrule
	\end{tabular}
\end{table}

\subsubsection{Dual-axis MIMO}
For the evaluation of the dual-axis MIMO system identification, we design two sets of combined inputs, each containing two excitation signals applied simultaneously to the X-axis and Y-axis. The first combination consists of a multi-frequency and multi-amplitude harmonic wave $u{{p}_{a1}}\left( t \right)=50+15\sin \left( 60\pi t \right)+25\sin \left( 100\pi t \right)$, and a simple sinusoidal signal $u{{p}_{a2}}\left( t \right)=50+40\sin \left( 80\pi t \right)$ for $t\in \left[ 0,200ms \right]$. The second combination consists of two excitation signals, $u{{p}_{b1}}(t)$ and $u{{p}_{b2}}(t)$, both composed of multiple cosine and sine waves with different phases, also lasting for 200ms. The sampling time step for all signals is $100\mu s$.

\begin{figure}[htbp!]
	\centering\includegraphics[width=1.0\columnwidth]{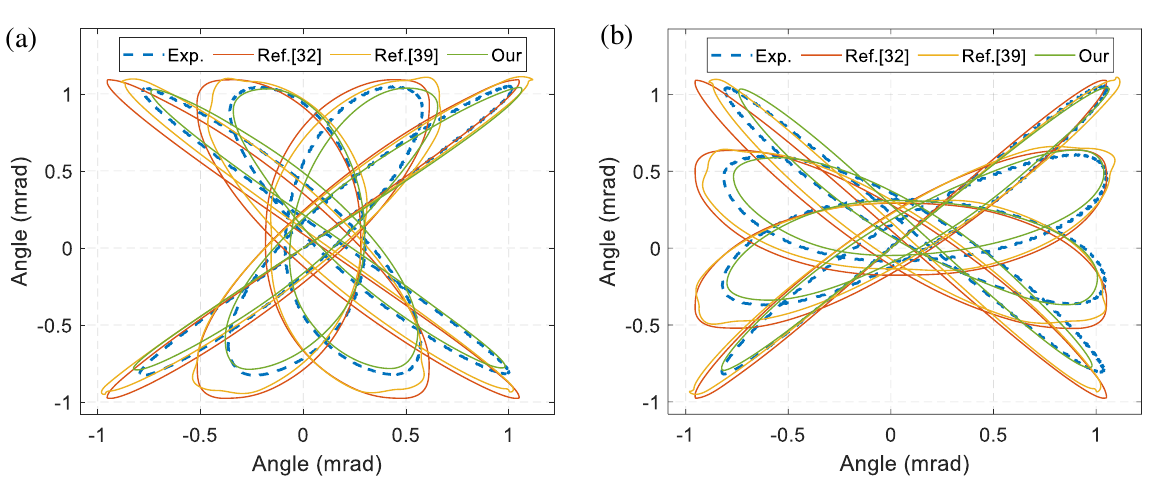}
	\caption{Comparison of the 2D trajectories of MIMO model outputs (solid lines) and experimental outputs (dashed lines) under dual-axis excitation with composite signals $u{{p}_{a1}}$ and $u{{p}_{a2}}$. (a) $u{{p}_{a1}}$ excites the X-axis while $u{{p}_{a2}}$ excites the Y-axis. (b) $u{{p}_{a2}}$ excites the X-axis while $u{{p}_{a1}}$ excites the Y-axis.}
	\label{fig:13}
\end{figure}

\begin{figure}[htbp!]
	\centering\includegraphics[width=1.0\columnwidth]{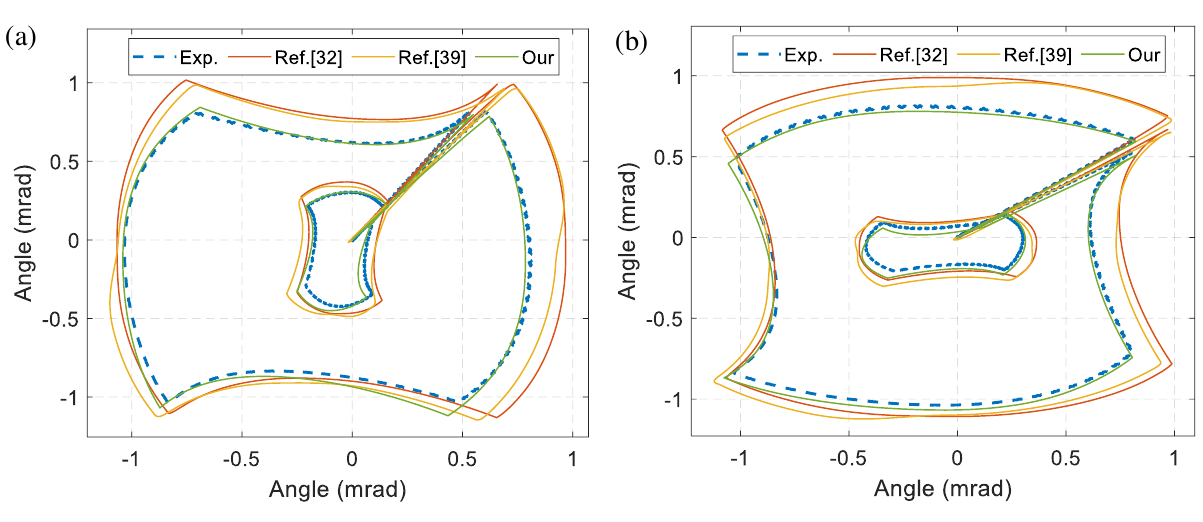}
	\caption{Comparison of the 2D trajectories for outputs of MIMO models (solid lines) and experiments (dashed lines) under dual-axis excitation with composite signals $u{{p}_{b1}}$ and $u{{p}_{b2}}$. (a) $u{{p}_{b1}}$ excites the X-axis while $u{{p}_{b2}}$ excites the Y-axis. (b) $u{{p}_{b2}}$ excites the X-axis while $u{{p}_{b1}}$ excites the Y-axis.}
	\label{fig:14}
\end{figure}

Under excitation with $u{{p}_{a1}}$ and $u{{p}_{a2}}$, the 2D trajectories for the outputs of MIMO models and experiments are shown in Fig. \ref{fig:13}(a) and (b). Similarly, the 2D trajectories with $u{{p}_{b1}}$ and $u{{p}_{b2}}$ are depicted in Fig. \ref{fig:14}(a) and (b). The proposed comprehensive model exhibits the best agreement between predicted outputs and actual trajectories. In contrast, model \cite{liu2019composite}, which does not consider cross-coupling, shows the most significant deviation from the predicted trajectories. Model \cite{wang2019comprehensive}, adopting a first-order inertial element to model cross-coupling, has limited accuracy, resulting in some deviation. Furthermore, the RMSE of the trajectories for the three comprehensive models under different excitations are listed in Tabs. 4 and 5. The average RMSE of the predicted trajectories using the proposed model is reduced by 71.2\% and 66.8\% compared to models \cite{liu2019composite} and \cite{wang2019comprehensive}, respectively. This reduction is more significant than that observed in the SISO system, reflecting the cumulative effects of hysteresis and cross-coupling identifications. These results demonstrate that the MIMO system dynamics based on the proposed comprehensive model can accurately describe the dual-axis PFSM system, thus validating the correctness of the system modeling and the effectiveness of the identification strategy.

\begin{table}[htpb!]
	\centering
	\caption{RMSE of comprehensive MIMO models under composite signal excitation}
	\label{tab:4}
	\begin{tabular}{@{}cllll@{}}
		\toprule
		& \multicolumn{1}{c}{$\left( u{{p}_{a1}},u{{p}_{a2}} \right)$} & \multicolumn{1}{c}{$\left( u{{p}_{a2}},u{{p}_{a1}} \right)$} & \multicolumn{1}{c}{$\left( u{{p}_{b1}},u{{p}_{b2}} \right)$} & \multicolumn{1}{c}{$\left( u{{p}_{b2}},u{{p}_{b1}} \right)$} \\ \midrule
		Model \cite{liu2019composite} & 0.166                 & 0.1601                & 0.1872                & 0.1842                \\
		Model \cite{wang2019comprehensive} & 0.1486                & 0.1427                & 0.1734                & 0.1739                \\
		Proposed Model & 0.0519                & 0.0444                & 0.0684                & 0.059                 \\ \bottomrule
	\end{tabular}
\end{table}

\section{Conclusion}\label{sec:5}
This paper introduces a novel high-precision open-loop model for PFSM, incorporating hysteresis, creep, mechanical, and cross-coupling dynamics with clear physical significance. Utilizing a Hammerstein structure, it effectively describes the nonlinear and rate-dependent characteristics of the PFSM system. Specifically, by augmenting the classic Bouc-Wen model, it remains rate-independent while evolving into a nonlinear asymmetric hysteresis model that can depict the phenomenon where PEA resistance to deformation is stronger at higher drive voltage.

For the linear rate-dependent component, the physical process from PEA displacement to PFSM angle is meticulously considered, and the modeling of mechanical cross-coupling between the dual axes is refined from first principles. Moreover, based on the isolation of dynamics of various modules at different frequency scales, appropriate excitation signals are chosen to minimize the interaction between dynamic components, and model parameters are identified step by step. The identification results of hysteresis, creep, and mechanical dynamics aligns well with the experimental measurements.

Composite excitation signals are applied simultaneously to both axes, the RMSE of the MIMO model prediction trajectories for the X-axis, Y-axis, and dual-axis are 0.44\%, 1.11\%, and 1.86\%, respectively, confirming the effectiveness of the proposed modeling and identification methods.

The proposed modeling approach can be employed to design various model-based high-precision PFSM controllers, which have significant value in applications such as free-space optical communication, stabilized imaging, and laser machining in beam precise pointing systems.

\section*{Disclosures}
The authors declare no conflicts of interest.

\bibliographystyle{unsrtnat}
\bibliography{references}  %%% Uncomment this line and comment out the ``thebibliography'' section below to use the external .bib file (using bibtex) .




% 	\bibitem{keshet2016prediction}
% 	Keshet, Renato, Alina Maor, and George Kour.
% 	\newblock Prediction-Based, Prioritized Market-Share Insight Extraction.
% 	\newblock In {\em Advanced Data Mining and Applications (ADMA), 2016 12th International 
%                       Conference of}, pages 81--94,2016.

% \end{thebibliography}

\end{document}